 \def\Version{June 28, 2004}
\documentclass{ws-ijmpd}
\usepackage{overcite}

\usepackage{epsf}
\usepackage{latexsym}

\def\beq{\begin{equation}}
\def\eeq{\end{equation}}

\def\dualp#1{{}^{\ast_{(\hbox{$\scriptstyle #1$})}} \kern-1pt}

\catcode`@=11
\def\meqalign#1{\null\,\vcenter{\openup\jot\m@th
  \ialign{\strut\hfil$\displaystyle{##}$&&$\displaystyle{{}##}$\hfil
      \crcr#1\crcr}}\,}
\catcode`@=12

\def\rmd{{\rm d}}
\def\rmD{{\rm D}}

\def\pmb#1{\setbox0=\hbox{$#1$}%
  \kern-.025em\copy0\kern-\wd0
  \kern.05em\copy0\kern-\wd0
  \kern-.025em\raise.0433em\box0}

\def\rmd{{\rm d}}
\def\fl{}

\def\vecA{\vec a(n)} \def\A{a(n)}

\begin{document}

%

\def\nocropmarks{\vskip5pt\phantom{cropmarks}}

\let\trimmarks\nocropmarks      

%

\markboth{Bini D., Cherubini C., Cruciani G., Jantzen R.T.}
{Geometric transport along circular orbits in stationary axisymmetric spacetimes
}

%
\catchline{}{}{}
%

\title{GEOMETRIC TRANSPORT ALONG CIRCULAR ORBITS IN STATIONARY AXISYMMETRIC SPACETIMES
}

\author{\footnotesize DONATO BINI}

\address{Istituto per le Applicazioni del Calcolo \lq\lq M. Picone\rq\rq, C.N.R.,
   I--00161 Roma, Italy \\
Sezione INFN di Firenze Polo Scientifico Via Sansone 1,
I--50019 Sesto Fiorentino (FI), Italy\\
International Centre for Relativistic Astrophysics, University of Rome, I--00185, Rome, Italy
\footnote{binid@icra.it}
}

\author{CHRISTIAN CHERUBINI}

\address{
Faculty of Engineering, University Campus Bio-Medico of Rome, via E. Longoni
47, I--00155 Rome, Italy\\
Institute of Cosmology and Gravitation, University of Portsmouth,
Portsmouth, PO1 2EG, UK\\
International Centre for Relativistic Astrophysics, University of Rome, I--00185, Rome, Italy
\footnote{cherubini@icra.it}
}

\author{GIANLUCA CRUCIANI}

\address{International Centre for Relativistic Astrophysics, University of Rome, I--00185, Rome, Italy
\footnote{cruciani@icra.it}
}

\author{ROBERT T. JANTZEN}

\address{Department of Mathematical Sciences, Villanova University,
  Villanova, PA 19085, USA
\\
International Centre for Relativistic Astrophysics, University of Rome, I--00185, Rome, Italy
\footnote{robert.jantzen@villanova.edu}
}

\maketitle

\maketitle

\begin{history}
\received{\Version}
\revised{Day Month Year}
\end{history}

\begin{abstract}
Parallel transport along circular orbits in orthogonally transitive stationary axisymmetric spacetimes is described explicitly relative to Lie transport in terms of the electric and magnetic parts of the induced connection. The influence of both the gravitoelectromagnetic fields associated with the zero angular momentum observers and of the Frenet-Serret parameters of these orbits as a function of their angular velocity is seen on the behavior of parallel transport through its representation as a parameter-dependent Lorentz transformation between these two inner-product preserving transports which is generated by the induced connection. This extends the analysis of parallel transport in the equatorial plane of the Kerr spacetime to the entire spacetime outside the black hole horizon, and helps give an intuitive picture of how competing ``central attraction forces" and centripetal accelerations contribute with gravitomagnetic effects to explain the behavior of the 4-acceleration of circular orbits in that spacetime.

\end{abstract}

\section{Introduction}

Orthogonally transitive stationary axisymmetric spacetimes play a central role in our ideas about rotation and angular momentum in the context of general relativity. Black hole spacetimes, the G\"odel spacetime, and even Minkowski spacetime expressed in coordinates adapted to this symmetry have shown how interesting and different
rotation can be in the relativistic setting compared to our intuition from nonrelativistic mechanics and Newtonian gravitation. The constant speed circular orbits in these spacetimes are crucial in almost every coordinate system used to represent them, and their geometry  reveals exactly how relativistic effects 
entangle
matters associated with rotating observers.

A family of test observers filling a region of spacetime and each individually rotating at a constant angular velocity (a circular observer family)
is described by a ``quasi-Killing" congruence \cite{iyevis}, namely a congruence associated with a vector field which is a stationary axially symmetric position-dependent linear combination of the two Killing vector fields describing the stationary axial symmetry of the spacetime. For spacetimes which are asymptotically flat away from the symmetry axis, two privileged families of observers exist \cite{greschvis}:
1) the static observers, or distantly nonrotating observers, following the trajectories of the timelike Killing vector field which is regular in that limit but, in general, is characterized by a nonzero vorticity;
2) the zero-angular-momentum observers (ZAMOs), or locally nonrotating observers, which have zero vorticity and are orthogonal to the axisymmetry Killing vector which has closed circular orbits.
The congruence of static observers determines the time coordinate lines while the family of orthogonal hypersurfaces of the ZAMOs determines the time coordinate hypersurfaces of the natural Boyer-Lindquist-like coordinate system so often used in representing these spacetimes \cite{mtw}.

The world lines of the ZAMOs and static observers
respectively have zero values of the angular momentum ``$p_\phi$" associated with the axisymmetry Killing vector field and of the ``angular velocity" $\dot\phi$ associated with the Killing vector field of the stationarity which is regular at spatial infinity away from the axis of symmetry, revealing two different aspects of the ``nonrotating" property for orbits in the nonrelativistic theory which become distinct from each other in the relativistic theory  when  the spacetime  itself is ``rotating."
In contrast to these two ``extrinsic scalar quantities" for circular orbits which are defined in relation to these two Killing fields, one also has two intrinsic scalar quantities whose extreme values characterize another pair of ``nonrotating" properties for a world line  in
the nonrelativistic theory, but which also become distinct in the relativistic theory in a rotating spacetime.
These intrinsic scalars are the Frenet-Serret curvature (magnitude of the 4-acceleration) and the magnitude of the Frenet-Serret angular velocity \cite{iyevis}.

At large distances the extremely accelerated observers
\cite{def95,pag98,sem98}
in some vague sense require the maximum outward acceleration to resist falling into a rotating central mass  since they have the least centrifugal force needed to counteract the attraction, while the minimum intrinsic rotation observers (MIROs)
\cite{bjdf0,circfs,bjdf}
are rotating the least in the sense that their Frenet-Serret frame,
which is in turn rigidly related to any stationary axisymmetric frame for circular orbits, rotates the least with respect to local gyro-fixed axes among all circular orbits of different angular velocities at a given radius and angular location.
Thus the usual nonrotating stationary circular orbits of nonrelativistic Newtonian theory (and of the static relativistic case) separate into 4 different kinds of orbits in the relativistic case of a rotating (stationary but not static) spacetime, each of which embodies a different aspect of the properties we associate with nonrotating motion in nonrelativistic theory. The MIROs turn out to be important in describing the geometry
of the acceleration and intrinsic rotation of the family of circular orbits, discussed in appendix A.

The differential properties of a spacetime with symmetry, arising from the  action of the symmetry group on the metric, are bound together along a Killing trajectory by a continuous family of Lorentz transformations of the tangent space along it which relates parallel transport to Lie transport, since both transports preserve inner products.
Given a frame of vector fields invariant under the symmetry group action, the induced connection matrix along a Killing trajectory, namely the value of the matrix-valued connection 1-form on the tangent vector to the curve, is antisymmetric when index-lowered with the metric reflecting the invariance of inner products,
and generates
this family of Lorentz transformations, modulo a constant linear transformation required to orthonormalize the frame.
Killing trajectories play a key role in interpreting the geometry of spacetimes which admit them and the geometrical properties of such curves associated with parallel and Lie transport are completely determined by this map, so it is of interest to understand its geometry.

In particular, much has been written about stationary axisymmetric spacetimes and especially black hole spacetimes because of their importance in describing physical systems and as models for understanding how rotation works in general relativity. Circular orbits in these spacetimes are Killing trajectories which are either helices or closed circles in the intrinsically flat world tube cylinders of the symmetry orbits. These are the closest one can come to the Newtonian circular orbits in which our experience is grounded and have generated a great deal of interest over the past half century since the G\"odel and Kerr solutions were found. While the usefulness of the literature on this topic is somewhat uneven, nevertheless its richness still offers interesting material to discuss, and the present problem seems worthy of exposition.

The Serret-Frenet approach \cite{iyevis,synge2,circfs} describes parallel transport along these Killing trajectory curves (when nonnull) representing constant speed circular orbits using a natural frame adapted to the local splitting of spacetime determined by their unit tangent vector $U$. On the other hand the language of gravitoelectromagnetism \cite{mfg,idcf1,idcf2} describes parallel transport in terms of the local splitting of spacetime associated with the 4-velocity $u$ of either the ZAMOs or the static observers, both of whose world lines belong to the same larger family of all stationary circular orbits. The relationship between these descriptions involves the transformations of gravitoelectromagnetism, which show how the kinematical decomposition of the complete connection along the observer family orbits adapted to their local splitting enters into the decomposition of the induced connection along a general orbit adapted to its own local splitting. In fact the whole electromagnetic analogy comes out of the decomposition of the covariant derivative of just the 4-velocity of a world line with respect to a family of observers. Here we extend the splitting to the entire covariant derivative along the world line for the case of transverse relative acceleration \cite{rok} that occurs for circular orbits, showing how the gravitoelectric, gravitomagnetic and space curvature fields affect general parallel transport.

A previous article \cite{bjm2}, following up work by Rothman et al \cite{rot} and by Maartens et al \cite{MMM}, investigated parallel transport in stationary axisymmetric spacetimes along circular orbits which are confined to a plane containing oppositely-rotating circular geodesics, like the equatorial plane of the Kerr spacetime or when additional cylindrical symmetry is present without translational motion in the additional symmetry direction. This allows a study of clock effects \cite{bjm01}, i.e., comparing periods of counter-revolving orbits defined in various ways, as well as circular holonomy. Parallel transport of an initial vector along such an orbit leads to a variable rotation or boost in an invariant stationary axisymmetric 2-plane relative to the corresponding Lie dragged vector. The equatorial plane of the Taub-Nut spacetime \cite{bcj02} has more general behavior that is typical of the general case studied here, with both a boost and simultaneous rotation in a pair of mutually orthogonal invariant stationary axisymmetric 2-planes relative to the corresponding Lie dragged vector. The present article examines the geometry of parallel transport along (in general accelerated) nonnull stationary circular orbits in general orthogonally-transitive stationary axisymmetric spacetimes, appropriate to describe such orbits off the equatorial plane in the Kerr spacetime. Parallel transport then generically leads to a more general ``Lorentz 4-screw" Lorentz transformation \cite{synge} generated by a 2-form whose interpretation in terms of electromagnetism helps put it into its canonical form corresponding to aligned electric and magnetic fields, a relationship not mentioned in current textbooks on relativity.

The Kerr spacetime, originally studied very briefly in this context
\cite{bollgiamtiom}, is used as an explicit example to illustrate these ideas.
By constructing the explicit parallel transport transformation, the study of the Serret-Frenet properties of circular orbits begun in [\citen{circfs}] is completed,
and some some additional light is shed on the comparable strengths of the centripetal, gravitoelectric and gravitomagnetic effects along them.

\section{Induced connection on Killing trajectories}

The term ``circular orbit" as used here refers to a Killing trajectory in a stationary axisymmetric spacetime, i.e., a parametrized curve whose tangent vector is a constant linear combination of the two independent Killing vector fields generating the symmetry group. They may also be characterized as integral curves of Killing vector fields. General considerations as described in this section apply to any kind of symmetry group acting as isometries on spacetime.

Let $e_\alpha$ ($\alpha=0,1,2,3$) be a frame which is invariant under the symmetry group of a spacetime with metric $g$ (signature $-+++$), and let $\omega^\alpha$ be its dual frame. The connection components and connection 1-form matrix are
\beq
  \nabla_{e_\alpha} e_\beta = \Gamma^\gamma{}_{\alpha\beta}\,e_\gamma\ ,\
  \omega^\alpha{}_\beta = \Gamma^\alpha{}_{\gamma\beta}\,\omega^\gamma\ .
\eeq

Given a parametrization of the Killing trajectory
$x^\alpha(\lambda)=x^\alpha(c(\lambda))$ and the corresponding tangent vector
$c'(\lambda)^\alpha = \rmd x^\alpha(\lambda)/\rmd\lambda$, a family of Lorentz transformations is generated by a fixed element of the Lie algebra of the Lorentz group which arises from the induced connection along the trajectory, namely the value of the connection 1-form matrix on the tangent vector of the parametrized curve, which is also Lie dragged along the curve (i.e., has a constant component matrix).
If $X$ is a tangent vector undergoing parallel transport along the curve, then its components in the invariant frame satisfy the constant coefficient system of linear first order differential equations
\beq
0 = \frac{\rmD X^\alpha}{\rmd\lambda}
  = \frac{\rmd X^\alpha}{\rmd\lambda} + \omega(c'(\lambda))^\alpha{}_\beta\,X^\beta
\ ,
\eeq
or
\beq\label{eq:defA}
  \frac{\rmd X^\alpha}{\rmd\lambda} = A^\alpha{}_\beta\,X^\beta\ ,\
  A^\alpha{}_\beta =-\omega(c'(\lambda))^\alpha{}_\beta =
   -\Gamma^\alpha{}_{\gamma\beta}\, \rmd x^\gamma/\rmd\lambda\ .
\eeq
Since inner products are preserved by parallel transport, the induced connection matrix is antisymmetric when index-lowered to its fully covariant form
\beq
  0= \frac{\rmD}{\rmd \lambda} (e_\alpha \cdot e_\beta) = -2 \, A_{(\alpha\beta)}
\ . \eeq Thus if one thinks of the constant matrix
$A=(A^\alpha{}_\beta)$ as a mixed-tensor $A=A^\alpha{}_\beta\,
e_\alpha \otimes \omega^\beta$ on the tangent space, this
condition makes its fully covariant form $A^\flat =
A_{\alpha\beta}\, \omega^\alpha \otimes \omega^\beta = \frac12
A_{\alpha\beta}\, \omega^\alpha \wedge \omega^\beta$ a 2-form and
places the mixed tensor in the Lie algebra of the action of the
Lorentz group of each tangent space along the curve. In an
orthonormal invariant frame, the matrix then belongs to the matrix
Lie algebra of the Lorentz group. Furthermore if one performs any
change of frame $e_\alpha\to e_{\alpha'}={(M^{-1})^\beta}_\alpha
e_\beta$ where $dM^\beta{}_\alpha/d\lambda =0$ holds along the
Killing trajectory, then in the new matrix of induced connection,
the coefficients are simply the new components of this mixed
tensor $A$ since there are no additional inhomogeneous terms in
the transformation law for the induced connection along the curve.
This is true for any new stationary axisymmetric frame along a
circular orbit in the context of the class of spacetimes of
interest.

A vector which is Lie transported along such a curve has constant
components with respect to the symmetry-invariant frame, so its
covariant derivative is then \beq\label{eq:lietransport}
\frac{\rmD X^\alpha}{\rmd\lambda} = -A^\alpha{}_\beta \,X^\beta\ .
\eeq Eigenvectors of the matrix $A^\alpha{}_\beta$ with eigenvalue
zero therefore correspond to parallel transported fields along the
curve which are values of a single symmetry-invariant vector
field. When the tangent to the curve itself is such an
eigenvector, the curve is a geodesic.

The parallel transport system of equations has the exponential solution
\beq\label{solution}
  X^\alpha(\lambda) = [e^{\lambda A}]^\alpha{}_\beta\, X^\beta(0)
\eeq
representing a family of Lorentz transformations (modulo a constant linear transformation which orthonormalizes the frame).
The nature of this family depends on the eigenvalue properties of the matrix generator $A$, which can only fall into one
among four possible cases that can be classified by using the invariants of $A$ under linear transformations.
This task is performed by first introducing the decomposition of its index-lowered 2-form $A^\flat$ into electric and magnetic parts with respect to any future-pointing timelike unit vector $u$ (representing the 4-velocity of an observer)
\beq\label{eq:AuEB}
 A^\flat = u^\flat \wedge \mathcal{E}(u) + \dualp{u} \mathcal{B}(u) \ ,
\eeq
where $\dualp{u} \mathcal{B}(u)$ is the spatial dual in the local rest space orthogonal to $u$ and $\mathcal{E}(u)$, $\mathcal{B}(u)$ are both 1-forms. While this decomposition is observer-dependent, under a change of observer these two fields undergo an electromagnetic boost under which the well-known quantities
\beq
   I_1 = \mathcal{E}(u) \cdot \mathcal{B}(u)
       =  {}^\ast\, \kern-1pt A_{\alpha\beta}\,A^{\alpha\beta}/4\ ,\
   I_2 =  ||\mathcal{E}(u)||^2  - ||\mathcal{B}(u)||^2
       = -  A_{\alpha\beta}\,A^{\alpha\beta}/2
\eeq
are Lorentz invariants related to the obvious matrix notation invariants through the relations
${\rm det}(A)=-I_1^2$ and ${\rm Tr}(A^2)=2\,I_2$.
It is convenient to refer to curves as electric-dominated or boost-dominated if $I_2>0$ and magnetic-dominated or rotation-dominated if $I_2<0$, separated by the case $I_2=0$.

In the terminology of Synge \cite{synge,bjm2}, either $A$ is
nonsingular with
$I_1\neq0$ (with four distinct nonvanishing eigenvalues) and a nonvanishing determinant
(so $A$ is also nonsingular in the ordinary matrix sense), or
semi-singular with
$I_1=0$ and $I_2 \neq0$
or singular
with $I_1=0$ and $I_2=0$, the latter two cases having a vanishing determinant (so $A$ is also singular in the ordinary matrix sense). The first case which is general corresponds to the generator of a simultaneous boost and rotation in a pair of mutually orthogonal 2-planes, the second case to the generator of a pure boost or rotation in a single 2-plane whose orthogonal 2-plane is invariant, and the last case to a null rotation when $A$ is nonzero. The existence of a parallel transported vector (including geodesics, where the tangent vector is parallel transported) requires $A$ to have a zero eigenvalue and so can only occur if $A$ is at least semi-singular.

For a nonnull Killing trajectory one can introduce the unit tangent vector
$U= c'(\lambda)/|c'(\lambda)\cdot c'(\lambda) |^{1/2}$, and use an arclength parametrization $\lambda=s$, so
$U^\alpha =\rmd x^\alpha/\rmd s$. One then has the decomposition
\beq\label{eq:AUEB}
 A^\flat = U^\flat \wedge \mathcal{E}(U) + \dualp{U} \mathcal{B}(U) \ ,
\eeq
where the index-raised vector $\mathcal{E}(U)^\sharp =-a(U)=-\rmD U/\rmd s$ is the sign-reversed second (covariant) derivative of the paramet\-rized curve. Its magnitude
$\kappa=||a(U)||\geq0$ is the Frenet-Serret curvature, while $\mathcal{B}(U)^\sharp$ is the Frenet-Serret angular velocity vector.
To avoid complicating the discussion with additional signs, consider only timelike or spacelike 4-velocities $U$ satisfying $u\cdot U\leq0$ (on the future side of the local rest space $LRS_u$ of $u$ within the whole tangent space), which can be parametrized by the relative velocity $\nu$ with respect to the observer $u$
\beq\fl\qquad
  U = \gamma ( u + \nu\,\hat e ) \ ,\
\gamma = |1-\nu^2|^{1/2}\ ,\
\nu\neq\pm1
\eeq
where $\hat e$ is a unit spacelike vector orthogonal to $u$,
or in the corresponding null case, an unnormalized tangent
\beq
 \nu=\pm1:  P_\pm = u \pm \hat e\ .
\eeq The limit
$[\nu,\,\gamma\,\nu,\,\gamma]\to[\pm\infty,\,\pm1,0]$ yields the
Killing trajectories orthogonal to $u$. When $U$ is timelike and
future-pointing, then the electric and magnetic parts of the
induced connection undergo a boost in the plane of $u$ and $U$
from (\ref{eq:AuEB}) to (\ref{eq:AUEB}) exactly like the electric
and magnetic fields in electromagnetism \cite{mfg}. The explicit
construction of the exponential solution (\ref{solution}) of the
parallel transport equations requires transforming $A$ to a
canonical form which allows the exponentiation to be performed
explicitly, the canonical form being block-diagonal in the case of
a non-null curve (closely related to the eigenvector
diagonalization), which is equivalent to Lorentz transforming a
nonnull electromagnetic field to a frame in which the electric and
magnetic fields are parallel.

\section{Circular orbits in a stationary axisymmetric spacetime}

Now specialize the discussion to stationary circular orbits in an
orthogonally-transitive stationary axisymmetric spacetime. In
Boyer-Lindquist-like (symmetry-adapted) coordinates
$(x^0,\,x^1,\,x^2,\,x^3)=(t,\,r,\,\theta,\,\phi)$ the spacetime
line element may be expressed in the form \beq \rmd s^2 =  \rmd
s^2_{(t,\phi)} +\rmd s^2_{(r,\theta)} , \eeq where
\begin{eqnarray}\label{eq:otsa}
\rmd s^2_{(t,\phi)} &=& g_{tt}\,\rmd t^2 +2g_{t\phi}\,\rmd t\,
\rmd \phi +g_{\phi\phi}\,\rmd \phi^2
\ ,\nonumber\\
\rmd s^2_{(r,\theta)} &=& g_{rr}\,\rmd r^2+g_{\theta\theta}\,\rmd
\theta^2 \ ,
\end{eqnarray}
and the metric coefficients do not depend on the coordinates $t$
and $\phi$. Let $e_{\hat i} = g_{ii}^{-1/2}\partial_i$
$(i=r,\,\theta,\,\phi)$ be the orthonormalized spatial frame
tangent to the constant time coordinate hypersurfaces. Orthogonal
transitivity is just the condition that the directions along the
orbits, namely the $t$-$\phi$ 2-plane in the tangent space, are
orthogonal to the $r$-$\theta$ 2-plane in the tangent space. As
will be seen shortly, it is natural to call these two subspaces
the circular velocity plane and the circular acceleration plane
respectively.

The unit normal $e_{\hat0}=n$ to the time coordinate hypersurfaces
completes this spatial triad to a stationary axisymmetric
orthonormal spacetime frame $e_{\hat\alpha}$, and represents the
4-velocity of the family of ZAMOs, a family of accelerated
observers with 4-acceleration $a(n)$. They are locally nonrotating
in the sense that their vorticity $\omega(n)$ vanishes but they
have nonzero expansion tensor $\theta(n)$ (minus the extrinsic
curvature of the constant $t$ hypersurfaces) whose nonzero
components can be completely described by the expansion vector
$\theta_{\hat \phi}^\alpha = \theta(n)^\alpha{}_\beta\,
{e_{\hat\phi}}^\beta$ due to the form (\ref{eq:otsa}) of the
metric.

Introducing the lapse function $N=(-g^{tt})^{-1/2}$ and shift
vector field with its only nonvanishing component
$N^\phi=g_{t\phi}/g_{\phi\phi}$ one has \beq
n=\frac{1}{N}[\partial_t -N^\phi\,\partial_\phi]\ ,\quad
n^\flat=-N\,\rmd t \eeq and the metric itself is \beq
  \rmd s^2=-N^2 \rmd t^2 + g_{\phi\phi}\,(\rmd\phi + N^\phi \rmd t)^2
       + \rmd s^2_{(r,\theta)}\ .
\eeq The ZAMO kinematical quantities then have nonzero components
only in the $r$-$\theta$ 2-plane of the tangent space
\begin{eqnarray}
\vecA & = & (a(n)^{\hat r},\,a(n)^{\hat\theta})=(\partial(\ln
N)/\partial\hat r ,\,\partial(\ln N)/\partial\hat\theta)\
,\nonumber\\
\vec\theta_{\hat\phi} & = & \frac12\, \sqrt{g_{\phi\phi}}\,
(\partial N^\phi/\partial\hat r,\,\partial
N^\phi/\partial\hat\theta) \ ,
\end{eqnarray}
where it is convenient to adopt a 2-vector notation indicated by
an arrow for pairs of orthonormal frame components of vectors
belonging to this subspace. Since the expansion scalar (trace of
the expansion tensor) is therefore zero, the expansion tensor
coincides with the shear tensor. In the static limit $N^\phi
\to0$, the shear vector $\vec\theta_{\hat\phi}$ vanishes.

A general Killing trajectory is a stationary circular orbit with
constant relative velocity along the Killing angular direction
$e_{\hat\phi}$. The family of nonnull Killing vector unit tangents
$U$ of uniformly rotating nonnull orbits at a given radius can be
parametrized directly by the relative velocity $\nu$ or by the
rapidity $\alpha$ as follows
\begin{eqnarray}\label{eq:Ualpha}
   U &=& \gamma(\pm\, n + \nu\,e_{\hat\phi})
\\
     &=& \cases{ \pm\cosh\alpha \,\, n + \sinh\alpha \,\, e_{\hat\phi}\ ,
                       & timelike: $|\nu|<1,\ \nu = \tanh\alpha$\,,\cr
                 \phantom{\pm}\sinh\alpha \,\, n\pm \cosh\alpha \,\, e_{\hat\phi}\ ,
                       & spacelike: $|\nu|>1,\ \nu = \pm\coth\alpha$\,,\cr}
\nonumber
\end{eqnarray}
where $\gamma =|1-\nu^2|^{-1/2}$ and in the future-pointing ($+$) timelike case which
will be of primary interest here:  $\gamma=\cosh\alpha$.
Note that for either case above
\beq\label{eq:dnudalpha}
 \frac{d\nu}{d\alpha} = \cases{\gamma^{-2}\,, & $U$ timelike,\cr
                               \mp\gamma^{-2}\,, & $U$ spacelike.\cr}
\eeq The angular velocity $\zeta$ also parametrizes the
half-family $U\cdot n<0$  for $\zeta\in(-\infty,\infty)$ \beq U
=\Gamma\, [\partial_t +\zeta \partial_\phi ]\ ,\ \Gamma =
\gamma/N,\ \,\zeta=-N^\phi+\frac{N}{\sqrt{g_{\phi\phi}}}\,\nu \ .
\eeq It is convenient to refer to the cases $|\nu|<1$ and
$|\nu|>1$ respectively as subluminal and superluminal relative
velocities and extend the terminology of acceleration to the
superluminal case.

These orbits are all open helices in spacetime except for the
limiting case $\nu^{-1}\to0$ (or equivalently $\zeta^{-1}\to0$) of
closed $\phi$-coordinate circles. For the null cases $\nu=\pm1$,
one must give up normalizing the tangent vector, letting instead
$P_\pm = n \pm e_{\hat\phi}$. As long as $\zeta\neq 0$, one can
use $\lambda=\phi$ as a parameter along  each such orbit, as done
in [\citen{bjm2}] where loops in $\phi$ were relevant to clock
effects, or one can use the arclength $s$ (proper time $\tau_U$)
as a parameter in the nonnull (timelike) case, as is convenient
for the Frenet-Serret approach. To translate the induced
connection matrix defined on the circular orbit by
Eq.~(\ref{eq:defA}) from one parametrization to the other for
future-pointing timelike $U$ with $\zeta\neq0$, one needs the
chain rule \beq
  \frac{\rmD X}{\rmd\tau_U} = \frac{\rmd\phi}{\rmd\tau_U}\,\frac{\rmD X}{\rmd\phi}
                      = \zeta\,\Gamma\,\frac{\rmD X}{\rmd\phi}
\quad \hbox{where} \quad \zeta\,\Gamma = g^{-1/2}\,\gamma\,(\nu -
N^{\hat\phi}/N)\ , \eeq so that \beq\label{eq:phi2tau}
A(\tau_U\hbox{-parametrization}) = \zeta\,\Gamma\,
A(\phi\hbox{-parametrization})\ . \eeq The proper time (or
arclength) parametrization will be used exclusively in this
article in place of the $\phi$-parametrization used previously
\cite{bjm2}.

The family of future-pointing timelike 4-velocities is a branch of a hyperbola
in the relative observer plane of $n$ and $e_{\hat\phi}$, namely the $t$-$\phi$ 2-plane in the tangent space spanned by the Killing vector fields, equivalently the subspace tangent to the cylindrical orbits of the symmetry group, called the circular velocity plane above. Another hyperbola in this plane describes the spacelike orbits,  with tachyonic 4-velocities, while the common asymptotes themselves correspond to the two oppositely rotating circular photons where no invariant preferred parametrization exists to normalize the tangent vector.

\section{The acceleration plane}

The corresponding curve of acceleration vectors for the family of future-pointing timelike circular orbits
is instead a branch of a hyperbola \cite{bjdf0,circfs,bjdf} in the orthogonal
tangent plane of the nonignorable coordinates $r$ and $\theta$, called the circular acceleration plane above but hereafter called the acceleration plane for short
\beq
   a(U) = \frac{\rmD U}{\rmd s}
     = \kappa\,(\cos\chi\, e_{\hat r}
               + \sin\chi\, e_{\hat\theta} )
\leftrightarrow
  \vec a(U) =(a(U)^{\hat r},\,a(U)^{\hat\theta}) \ ,
\eeq
where $(\kappa,\chi)$ are polar coordinates in the
$e_{\hat r}$-$e_{\hat\theta}$ plane
in the tangent space, both constant for a given orbit.
The Frenet-Serret angular velocity traces out a branch of a complementary
hyperbola centered at the origin with the same asymptote orientations.
The vertex of both hyperbolas occurs at the same relative velocity
$\nu_{\rm(vert)}$
which characterizes the MIRO family.
This geometry of the acceleration plane is developed further in appendix A.
The second branches of these two hyperbolas correspond to the spacelike orbits, for which it is convenient to extend the use of the word acceleration to refer to $a(U)$, while the two asymptotes correspond to the acceleration vectors of the two oppositely rotating circular photon orbits.

The acceleration $a(U)$
of the family of future-pointing timelike circular orbits in a stationary axisymmetric spacetime can be expressed in terms of the ZAMO relative observer decomposition in the following form (see equation (6.1) of [\citen{circfs}], where the notation $\vec A$ is used for $\vecA$)
\begin{eqnarray}\label{eq:akthetaA}
  \vec a(U)
  &=&  \gamma^2\,[ \vec k_{(\rm lie)}\, \nu^2
      + 2\,\vec\theta_{\hat\phi}\, \nu
      + \vecA ]
  \nonumber\\
  &=&   \vec k_{(\rm lie)}\, \sinh^2\alpha
      + 2\,\vec\theta_{\hat\phi}\, \sinh\alpha \,\cosh\alpha
      + \vecA \cosh^2\alpha
  \nonumber\\
  &=& \frac12 (\vec k_{(\rm lie)} +\vecA)\cosh 2\alpha
  + \vec \theta_{\hat \phi}\,\sinh 2\alpha
  + \frac12 (-\vec k_{(\rm lie)} +\vecA)
  \nonumber\\
  &=&\frac14 (\vec a_+ +\vec a_-)\cosh 2\alpha
  + \frac14 (\vec a_+ -\vec a_-)\sinh 2\alpha
  + \vec a_0 \ ,
\label{acc}
\end{eqnarray}
where $\vec k_{(\rm lie)}$ are the components of the ZAMO  Lie
relative curvature vector $k_{({\rm lie},U,n)}$ (for more details,
see [\citen{idcf2,bjdf}]), and $\vec a_\pm = \vec k_{(\rm lie)}
+\vecA \pm 2\,\vec\theta_{\hat \phi}$ are the
corotating/counter-rotating photon circular orbit accelerations
(corresponding to a unit ZAMO energy affine parameter, ie., using
$P=n\pm e_{\hat\phi}$ in place of $U$) which are aligned with the
asymptotes of the acceleration hyperbola and $\vec a_0 =\frac12
(\vecA-\vec k_{(\rm lie)})$ is the center of the acceleration
hyperbola. Fig.~2 of [\citen{circfs}] shows some typical examples
of these hyperbolas for the Kerr spacetime.

\begin{table}[t]
\vbox{ \hrule\vskip 3pt Table 1. Geometric quantities for the Kerr
spacetime, using the abbreviations $\Delta=r^2-2\,m\,r+a^2$,
$\Lambda = (r^2+a^2)^2-a^2\,s^2\Delta$, $\Sigma = r^2+a^2\,c^2$,
$(c,\,s) = (\cos \theta,\,\sin \theta)$. The orthonormal
components of vectors in the $r$-$\theta$ plane are denoted by
$\vec X= (X^{\hat r}, X^{\hat \theta})$.

\vskip 3pt\hrule

\begin{eqnarray}
(N,\,N^\phi) &=& \left(
\left[\Delta\,\Sigma/\Lambda\right]^{1/2},\,
 -2\,a\,m\,r/\Lambda
\right)
\nonumber\\
(g_{rr},\,g_{\theta\theta},\,g_{\phi\phi})
 &=& (\Sigma/ \Delta,\,\Sigma,\,\Lambda\,s^2/\Sigma )
\nonumber
\end{eqnarray}
\begin{eqnarray}
\vecA & = & \frac{m}{\Lambda\,\Sigma^{3/2}} \left(
\left[(r^2-a^2c^2)(r^2+a^2)^2-4mr^3a^2s^2\right]\Delta^{-1/2},
\, -2\,r\,a^2 s\,c\,(r^2+a^2) \right)\nonumber\\
\vec\theta_{\hat \phi} & = & \frac{m}{\Lambda\,\Sigma^{3/2}}
\left(-a\,s\,\left[r^2\,(3\,r^2+a^2)+
a^2\,c^2\,(r^2-a^2)\right],\, 2\,r\,a^3\,s^2\,c\,\Delta^{1/2}
\right)
\nonumber\\
\vec k_{(\rm lie)} & = & \frac{1}{\Lambda\,\Sigma^{3/2}} \left(
-\Delta^{1/2}\left[r\,\Sigma^2 -
m\,a^2\,s^2\,(r^2-a^2\,c^2)\right], \right.\nonumber\\
&&\hspace{13.2mm} \left.
-\frac{c}{s}\,\left[(r^2+a^2)\,\Sigma^2+2\,m\,
r\,a^2\,s^2\,(r^2+a^2+\Sigma)\right]\right) \nonumber
\end{eqnarray}
\vskip3pt \hrule}
\end{table}

Table 1 shows the geometric quantities needed to analyze the acceleration plane geometry for the Kerr spacetime.
Discussion below will be limited to the case $0\leq a \leq m$ and to radii
outside the black hole outer horizon at $r_{\rm(hor)} = m+(m^2-a^2)^{1/2}$
where $\Delta=0$, with $a>0$ characterizing a black hole corotating in the positive $\phi$ direction.
In the equatorial plane $\theta=\pi/2$,
where parallel transport was discussed extensively in [\citen{bjm2}],
the only nonvanishing components are radial
\begin{eqnarray}
\A_{\hat r} & = &
\frac{m\,\Delta^{-1/2}\left[(r^2+a^2)^2-4\,a^2\,m\,r\right]}
{r^2\,(r^3+a^2\,r+2\,m\,a^2)}\nonumber\\
{\theta_{\hat \phi}}_{\hat r} & = &
\frac{m\,a\,(3\,r^2+a^2)}{r^2\,(r^3+a^2\,r+2\,m\,a^2)}\nonumber\\
{k_{(\rm lie)}}_{\hat r} & = &
\frac{\Delta^{1/2}(r^3-m\,a^2)}{r^2\,(r^3+a^2\,r+2\,m\,a^2)}\ .
\end{eqnarray}
The roots $\nu_+>0$ and $\nu_-<0$ of the quadratic expression in
$\nu$ in square brackets in the first expression
(\ref{eq:akthetaA}) for the purely radial acceleration in the
equatorial plane define the ZAMO relative velocities of the
corotating and counterrotating circular geodesics respectively
(assuming $a>0$) \beq
 \nu_\pm =\frac{r^2+a^2\mp 2\,a\,(m\,r)^{1/2}}{\Delta^{1/2}[a\pm r\,(r/m)^{1/2}]}\ ,
\eeq which are related to the quadratic form coefficients by \beq
\A_{\hat r} = {k_{(\rm lie)}}_{\hat r}\,\nu_+\nu_-\ , \quad
{\theta_{\hat \phi}}_{ \hat r} = -{k_{(\rm lie)}}_{\hat
r}\,(\nu_++\nu_-)/2\ . \eeq

In the context of spacetime splitting techniques
(``gravitoelectromagnetism" \cite{mfg,idcf1,idcf2}), one can mimic
the Newtonian situation by introducing a ``relative" gravitational
force as seen by the ZAMOs which is analogous to the Lorentz force
of electromagnetism. Since $n$ is a stationary vector field, its
intrinsic derivative along $U$ (see Eq.~(\ref{eq:lietransport}))
is \beq F^{(G)\,\alpha}_{\rm(fw)} \equiv -\frac{D
n^\alpha}{d\tau_U} = A^\alpha{}_\beta\,n^\alpha =
\mathcal{E}(n)^\alpha \eeq which lies in the acceleration plane
and in turn may be expressed in terms of the ZAMO
gravitoelectromagnetic vector fields by \beq \vec
F^{(G)}_{\rm(fw)} =\gamma\,[\vec g -\nu\,\vec \theta_{\hat \phi}]
=\gamma\,[\vec g +\nu\,e_{\hat \phi}\times_n \vec H]\ , \eeq where
$\vec g=-\vecA$ are the components of the gravitoelectric vector
field and $\vec H$ are the components of the effective
gravitomagnetic vector field $H=e_{\hat \phi}\times_n \theta_{\hat
\phi} $ due to the ``differential rotation" of the ZAMOs arising
from the shear of the ZAMO 4-velocity field $n$. The latter field
satisfies $\theta_{\hat \phi}=-e_{\hat \phi}\times_n H$ and
$||\theta_{\hat \phi}|| = ||H||$, where $\times_n$ is the natural
spatial cross-product in the local rest space of the ZAMOs.

Similarly the covariant derivative of the stationary vector field
$e_{\hat\phi}$, which defines the Fermi-Walker relative curvature
vector \cite{fsfw}, leads to \beq \frac{D
e_{\hat\phi}^\alpha}{d\tau_U} \equiv \gamma\,\nu\,
k_{\rm(fw)}^\alpha = - A^\alpha{}_\beta\,e_{\hat\phi}^\beta =
-{[\dualp{n} \mathcal{B}(n)]^\alpha}_\beta\;{e_{\hat\phi}}^\beta =
[\mathcal{B}(n)\times_n e_{\hat\phi}]^\alpha\ . \eeq Since
$\mathcal{B}(n)$ has no components along $e_{\hat\phi}$, this
inverts to \beq
 \mathcal{B}(n)^\sharp =  e_{\hat\phi} \times_n [\gamma\,\nu\, k_{\rm(fw)}]
\quad \hbox{or}\quad
  \dualp{n}\mathcal{B}(n) = \omega^{\hat\phi}\wedge [ \gamma\,\nu\, k_{\rm(fw)}]^\flat\ .
\eeq
One can interpret this as the Fermi-Walker relative angular
velocity vector associated with the changing direction of the
trajectory in the local rest space of the family of ZAMOs along
the world line (dividing out the gamma factor corresponds to ZAMO
proper time along the trajectory). Given these relationships, the
acceleration of the orbit can also be written in various ways
\begin{eqnarray}
a(U)
 &=&-\gamma^2 [g -2\,\nu\, \theta_{\hat \phi}  -\nu^2\, k_{(\rm lie)}  ]
    = - \gamma\, F^{(G)}_{\rm(fw)} + (\gamma\,\nu)^2\, k_{(\rm fw)}
\nonumber\\
 &=& -\gamma\,[\mathcal{E}(n)^\sharp + \nu\, e_{\hat\phi} \times_n\, \mathcal{B}(n)^\sharp]
\ .
\end{eqnarray}

The Fermi-Walker relative curvature is related to the Lie relative curvature for a stationary circular orbit by Eq.~(8.7) of [\citen{fsfw}]
\beq
k_{\rm (fw)}= k_{(\rm lie)} +\nu^{-1} \theta_{\hat \phi}\ .
\eeq
Using this to re-express $\mathcal{B}$ then leads to the final pair of equations
\begin{eqnarray}\label{eq:EB(n)}
 \mathcal{E}(n)^\sharp &=&  \gamma \, [g +\nu\, e_{\hat \phi}\times_n \,H]
                =  \gamma \, [g -\nu\, \theta_{\hat \phi}]
\ ,\nonumber\\
 \mathcal{B}(n)^\sharp &=& \gamma\,[ \nu \,e_{\hat\phi} \times_n \,k_{\rm(lie)} + H]
                = \gamma\, e_{\hat\phi} \times_n\, [\nu\, k_{\rm(lie)} + \theta_{\hat \phi}]
\ .
\end{eqnarray}
For the ZAMO orbits ($\nu=0$, $\gamma=1$), the electric part of $A$ is just the gravitoelectric vector field and the magnetic part of $A$ is just the effective gravitomagnetic vector field.

The relations (\ref{eq:EB(n)}) almost have the form of an electromagnetic boost except for the two distinct accelerations, but introducing the sum and difference fixes this leaving an extra correction. Letting $g_\pm = (g \pm k_{\rm(lie)})/2$, one finds
\beq\label{eq:EBM}
\pmatrix{\mathcal{E}(n)^\sharp \cr \mathcal{B}(n)^\sharp\cr}
=  M(\alpha) \pmatrix{g_-\cr  H\cr}
 + M(-\alpha) \pmatrix{g_+\cr  0\cr}\ ,
\eeq where \beq M(\alpha) = \pmatrix{\gamma & \gamma\,\nu\,
e_{\hat \phi}\times_n \cr
                     - \gamma\,\nu\, e_{\hat \phi}\times_n & \gamma\cr}
= \pmatrix{ \cosh \alpha &\sinh \alpha\,e_{\hat \phi}\times_n \cr
- \sinh \alpha\,e_{\hat \phi}\times_n & \cosh \alpha \cr } \eeq is
the boost operator for a pair of electric and magnetic fields in
the $r$-$\theta$ 2-plane $LRS_n \cap LRS_U$ orthogonal to the
direction of relative motion in the local rest space subspace
common to both $n$ and $U$, where the projection $P(n,U)^{-1}$ of
the general formula (4.14) of [\citen{mfg}] acts as the identity
transformation. This is true for world lines characterized by
purely transverse relative acceleration \cite{rok}, as are
circular orbits seen by circularly orbiting observer families. The
rapidity parametrization of the boost matrix here only holds when
$U$ is future-pointing and timelike; the remaining cases of
Eq.~(\ref{eq:Ualpha}) must be considered separately.

\section{Parallel transport along circular orbits and the Frenet-Serret description}

The index-lowered matrix governing parallel transport along an
arclength (proper time when timelike) parametrized circular orbit
with 4-velocity $U$ is \beq\label{matra} A_{\alpha\beta} =
-(\Gamma \,\zeta)\, [ g_{\phi[\alpha,\beta]}+ \zeta^{-1}\,
g_{t[\alpha,\beta]} ] =-U^\mu g_{\mu [\alpha,\beta]}\, \eeq where
the factor $ \Gamma \zeta={\gamma\, \zeta}/{N}=\rmd \phi /\rmd s$
takes into account the arclength / proper time parametrization of
the trajectory and would instead have the value 1 if the curve
were parametrized by the angle $\phi$ as in previous articles
\cite{bjm2,bcj02}. The 2-form $A^\flat$ can be expressed in terms
of its electric and magnetic parts either with respect to $U$ or
with respect to the ZAMOs
\begin{eqnarray}
\label{matrUn}
A^\flat = U^\flat \wedge \mathcal{E}(U) + \dualp{U} \mathcal{B}(U)
= n^\flat \wedge \mathcal{E}(n) + \dualp{n} \mathcal{B}(n)\ .
\end{eqnarray}

Because of the orthogonally-transitive stationary axisymmetric
form (\ref{eq:otsa}) of the metric components, the matrix
$(A^\alpha{}_\beta)$ can only have nonvanishing components for
index pairs $(t,r)$, $(t,\theta)$, $(\phi,r)$, $(\phi,\theta)$ in
either order. This means that its corresponding electric and
magnetic vector parts can only have nonzero components in the
$r$-$\theta$ plane, namely the acceleration plane where all the
acceleration terms lie. The transformation between the electric
and magnetic quantities relative to $U$ and those corresponding to
$n$ defines the electromagnetic boost already introduced above
\beq
 \pmatrix{ \mathcal{E}(U)^\sharp\cr \mathcal{B}(U)^\sharp\cr }
=\pmatrix{ \gamma\, [\mathcal{E}(n)^\sharp+\nu\, e_{\hat \phi}
\times_n\, \mathcal{B}(n)^\sharp]\cr \gamma\,
[\mathcal{B}(n)^\sharp-\nu\, e_{\hat \phi} \times_n\,
\mathcal{E}(n)^\sharp]\cr } = M(\alpha)
\pmatrix{\mathcal{E}(n)^\sharp \cr \mathcal{B}(n)^\sharp\cr} \ .
\eeq Thus this represents a transformation of electric and
magnetic fields in the acceleration plane as well. Composing this
with (\ref{eq:EBM}) leads to the fact that the $U$ fields only
involve the relative velocity through an inhomogeneous boost by
twice the rapidity of the first boost 
\beq \label{eq:EBUgH}
\pmatrix{ \mathcal{E}(U)^\sharp\cr \mathcal{B}(U)^\sharp\cr } =
M(2\,\alpha) \pmatrix{ g_- \cr H \cr} + \pmatrix{ g_+ \cr 0 \cr}\,
, 
\eeq 
recalling that $g_\pm =(g \pm k_{(\rm lie)})/2$. The extra
term for the electric part $\vec g_+=-\vec a_0$ is the sign-reversed
center of the acceleration hyperbola, representing the shift of
origin needed to follow the double boost which takes the
(modified) ZAMO gravitoelectric and gravitomagnetic vectors to the
electric and magnetic parts of $A$.

The decomposition of $A$ with respect to $U$ can be interpreted by expressing it in terms of the associated spacetime Frenet-Serret frame $\{e_0=U,e_1,e_2,e_3\}$ and the curvature and torsions $\kappa$, $\tau_1$ and $\tau_2$ \cite{iyevis,circfs}
\begin{equation}
\frac{D e_0}{d\tau_U} = \kappa\, e_1\ ,\ \frac{D e_1}{d\tau_U} =
\kappa\, e_0 +\tau_1\, e_2\ ,\ \frac{D e_2}{d\tau_U} = -\tau_1
\,e_1+ \tau_2 \,e_3\ ,\ \frac{D e_3}{d\tau_U} = -\tau_2\, e_2\ .
\end{equation}
From the identification
\beq
   \frac{D e_\alpha}{d\tau_U} = -A^\beta{}_\alpha\, e_\beta\ ,
\eeq
one can read off the identifications
\begin{eqnarray}\label{eq:EBtauomega}
A^\sharp &=&-\kappa\, U\wedge e_1 +\tau_1\, e_1\wedge e_2
+\tau_2\, e_2\wedge e_3
\ ,\nonumber\\
\mathcal{E}(U)^\sharp &=& -a(U) =  -\kappa\, e_1\ ,\quad
\mathcal{B}(U)^\sharp = \omega_{\rm (FS)} = \tau_1\, e_3 +\tau_2\,
e_1 \ ,
\end{eqnarray}
assuming that $\{e_1,\,e_2,\,e_3\}$ is a right-handed triad (so
that $e_3=e_1\times_U \,e_2$). Both $\mathcal{E}(U)^\sharp$ and
$\mathcal{B}(U)^\sharp$ belong to the acceleration plane
span($\{e_1,\,e_3\}$) = span($\{e_{\hat r},\,e_{\hat\theta}\}$),
while $\{e_0,\,e_2\}$ span the velocity plane. By continuity in
the rapidity $\alpha$, the Frenet-Serret frame may be easily
extended to geodesic orbits where $\kappa=0$.

The Frenet-Serret scalars completely describe parallel transport
along a circular orbit with respect to a frame determined by the
curvature properties of the orbit itself. Because the unit tangent
of the constant speed circular orbit is stationary, so too is its
entire Frenet-Serret frame, which on a given trajectory is related
to the natural symmetry adapted frame associated with the
Boyer-Linquistlike coordinates by a constant linear
transformation.

The Synge classification of the  matrix $A$ is easily translated
into conditions on the intrinsic Frenet-Serret quantities
\begin{eqnarray}\label{eq:syngekt}
I_1 &=& \mathcal{E}(U)\cdot \mathcal{B}(U)
     = - a(U) \cdot \omega_{\rm(FS)}
     = -\kappa\, \tau_2\ ,\nonumber\\
I_2 &=& ||\mathcal{E}(U)||^2-||\mathcal{B}(U)||^2
      = \kappa^2-||\omega_{\rm(FS)}||^2
      = \kappa^2 -(\tau_1^2+\tau_2^2)
    \ .
\end{eqnarray}
Thus $A$ is nonsingular ($I_1\neq 0$) in both the Synge sense and
the ordinary linear algebra matrix sense  if the orbit $U$ is
nongeodesic ($\kappa\neq0$) and has nonzero second torsion. $A$ is
semi-singular ($I_1=0$ but $I_2\neq0$) if either $U$ is geodesic
or the second torsion is zero, while it is singular ($I_1=0=I_2$)
if in addition Frenet-Serret angular velocity has the same
magnitude as the acceleration.

For orthogonally-transitive stationary axially symmetric
spacetimes, the two torsions are simply related to the polar form
of the acceleration vector in the acceleration plane under certain
assumptions about the choice of the directions of the
Frenet-Serret frame vectors \cite{circfs} as described in detail
in Appendix B. For the future-pointing timelike case, one has
\beq\label{eq:a_kappatau}
 a(U) = \kappa\,(\cos\chi\, e_{\hat r} + \sin\chi\, e_{\hat\theta})\ ,\
\tau_1 = -\frac12\, \frac{d \kappa}{d\alpha}\ ,\ \tau_2 =
-\frac12\, \kappa \,\frac{d \chi}{d\alpha}\ , \eeq which can be
converted back to $\nu$-derivatives using
Eq.~(\ref{eq:dnudalpha}). Vanishing first torsion corresponds to a
critical point of $\kappa$ as a function of the rapidity $\alpha$
(including the case of the extremely accelerated orbits), while
vanishing second torsion corresponds to a critical point of the
polar angle $\chi$.

Finally, note that the corotating and counterrotating circular photon orbit tangent vectors $P_\pm = n\pm e_{\hat\phi}$ are parallel transported along $U$ when
$ D(n   \pm  e_{\hat \phi} )/d\tau_U = 0$. This translates into the statement
\beq\label{eq:E=B}
\mathcal{E}(n)
= Dn/\tau_U = \pm De_{\hat \phi}/d\tau_U
= \pm \mathcal{B}(n) \times_n e_{\hat\phi}\ ,
\eeq
which implies that
$\mathcal{E}(n)$ and $\mathcal{B}(n)$ are orthogonal and equal in magnitude, so both invariants $I_1=0=I_2$ vanish corresponding to the Synge singular case.

\typeout{!!! forced page break here to force figure caption and full page figure out. correct in final version !!!}

\vfill
\begin{figure}[!hb]
\typeout{Figure 1 caption by itself at bottom of same even/odd
page pair as full page graphic} \caption{The Frenet-Serret scalars
plotted versus relative velocity $\nu$ at $r/m=4,\, 3,\, 2$ (top
to bottom) and $\theta=\pi/2,\, \pi/4$ (left to right) for
future-pointing timelike circular orbits in an $a/m=1/2$ Kerr
spacetime: $\kappa$ (solid), $\tau_1$ (short-dashed), $\tau_2$
(long-dashed, absent for $\theta=\pi/2$  where $\tau_2=0$) and
$||\omega_{\rm(FS)}||$ (dotted, just $|\tau_1|$ for
$\theta=\pi/2$). The radii chosen for these three figures belong
to the equatorial plane radial intervals where $a$) two
oppositely-rotating timelike geodesics exist, $b$) one corotating
timelike geodesic exists, and $c$) no timelike geodesics exist,
which for $\theta=\pi/4$ roughly correlate respectively with the
regions A, B, C in Fig.~1 of [\citen{circfs}] characterized by the
same number of timelike orbits with zero radial acceleration at a
given $(r,\theta)$ location. } \label{fig:1}
\end{figure}

\typeout{Figure 1 graphic full page}
\begin{figure}[!hp]
\typeout{*** EPS figures 1-1,1-2,1-3,1-4,1-5,1-6}
\centerline{
\vbox{
\hbox{
\parbox{6cm}{\hbox to 6cm{\footnotesize\hfil $a$) $r=4m>r_{\rm geo-},\qquad \theta=\pi/2$\hfil}
\epsfysize=6cm\epsfbox{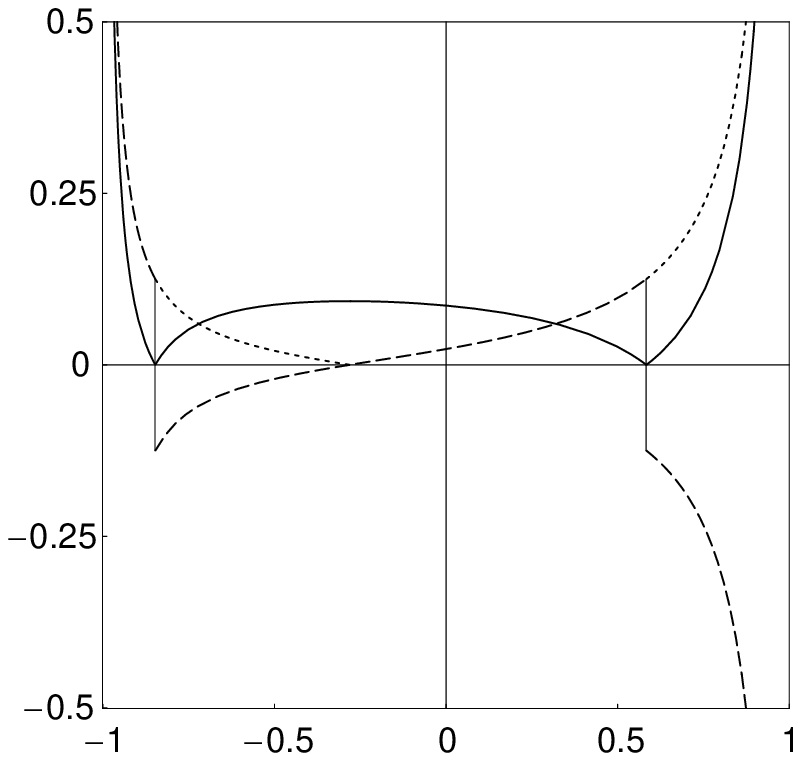}}
\hfill
\parbox{6cm}{\hbox to 6cm{\footnotesize\hfil \qquad$ \theta=\pi/4$\hfil}
\epsfysize=6cm\epsfbox{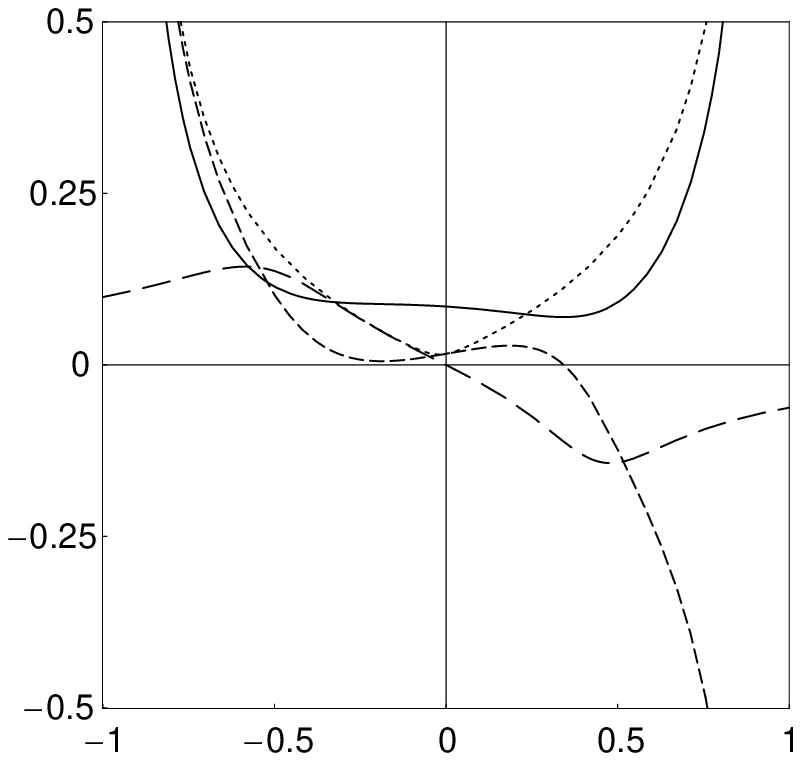}}
}
\hbox{
\parbox{6cm}{\hbox to 6cm{\footnotesize\hfil $b$) $r_{\rm geo+}<r=3m<r_{\rm geo-},\quad \theta=\pi/2$\quad}
\epsfysize=6cm\epsfbox{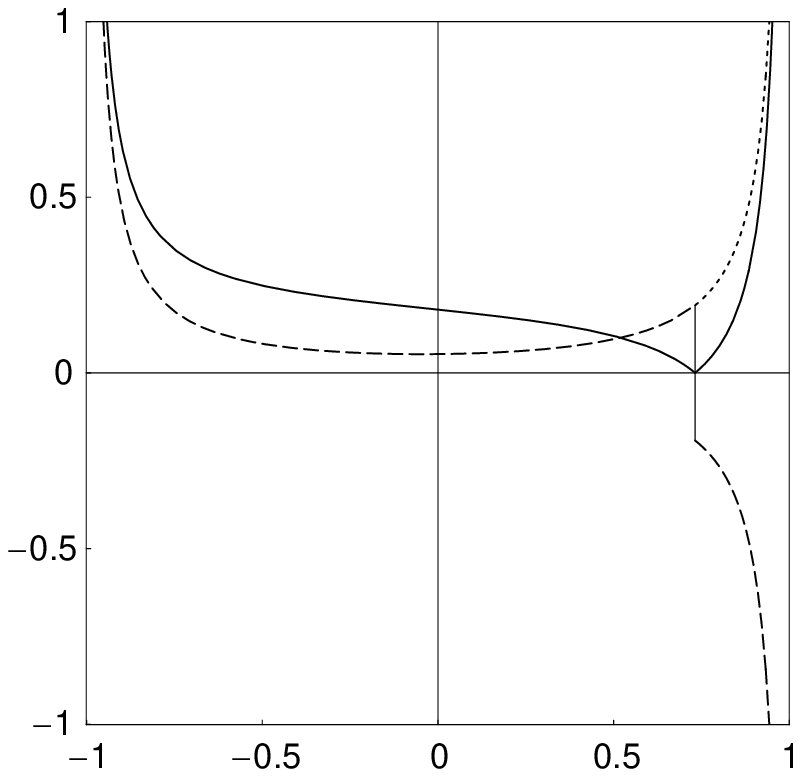}}
\hfill
\parbox{6cm}{\hbox to 6cm{\footnotesize\hfil \qquad$\theta=\pi/4$\hfil}
\epsfysize=6cm\epsfbox{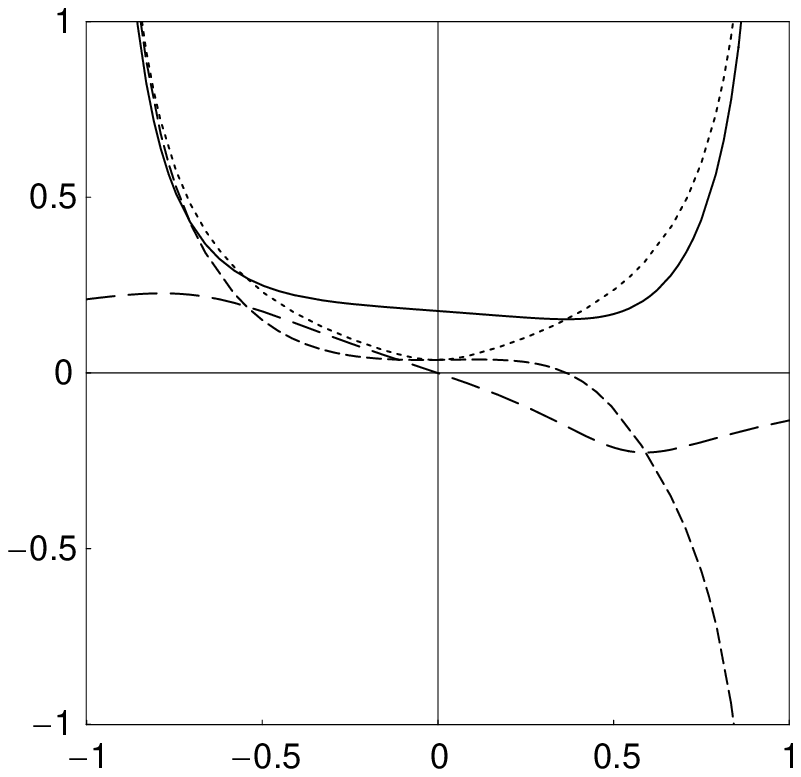}}
}
\hbox{
\parbox{6cm}{\hbox to 6cm{\footnotesize\qquad $c$) $r=2m<r_{\rm geo+},\qquad \theta=\pi/2$\hfil}
\epsfysize=6cm\epsfbox{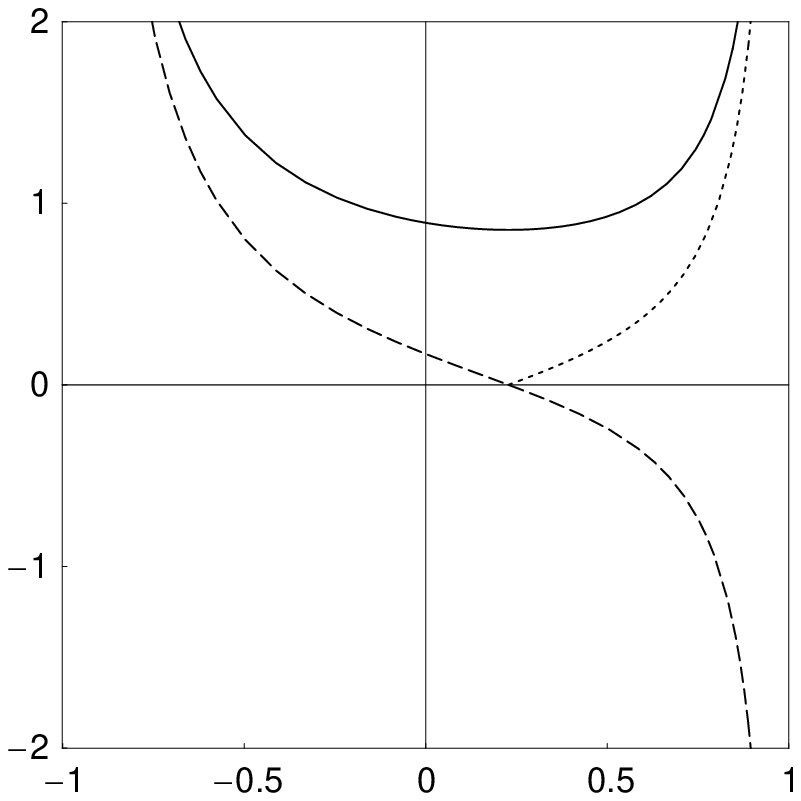}}
\hfill
\parbox{6cm}{\hbox to 6cm{\footnotesize\hfil \qquad$ \theta=\pi/4$\hfil}
\epsfysize=6cm\epsfbox{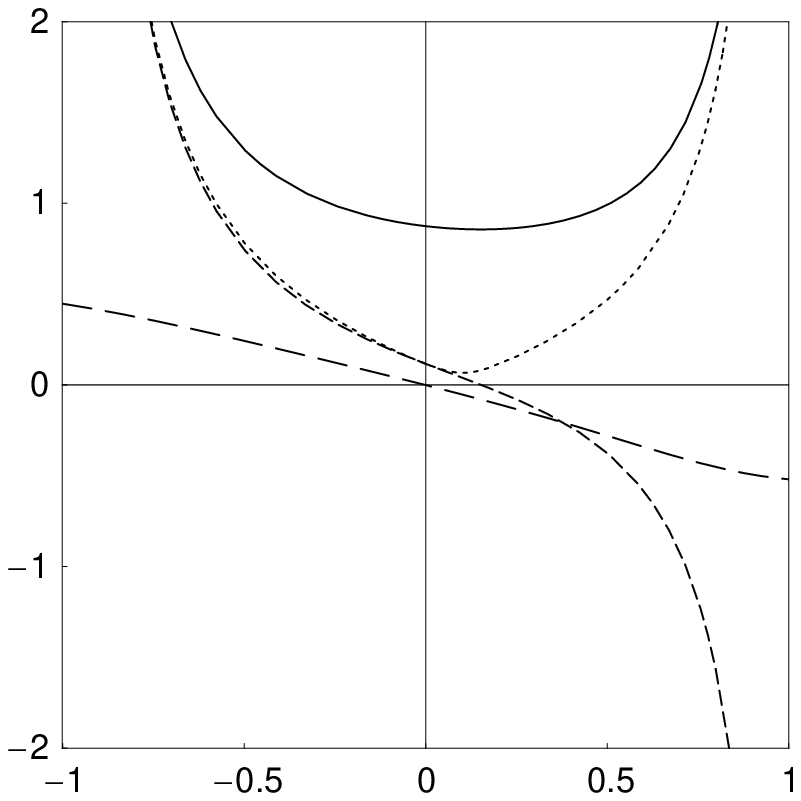}}
}
}
}
\end{figure}\clearpage

To make these abstractions more concrete in the Kerr spacetime, the Frenet-Serret frame has been defined so that its limit as $r\to\infty$ has the slow-motion far-field Schwarzschild properties of circular orbits: the acceleration is approximately radially outward
while the boosted direction of increasing $\phi$ is approximately along the $\phi$-coordinate direction, leading to
\beq
  e_1 \sim e_{\hat r}\ ,\
  e_2 \sim e_{\hat \phi}\ ,\
  e_3 = e_1 \times_U e_2 \sim - e_{\hat \theta}\ .
\eeq
The leading contribution to the first torsion is the special relativistic angular velocity
$\tau_1 \sim \nu/r$ of the rotating radial unit vector along a circular orbit of ZAMO relative velocity $\nu$. For a static spacetime this remains an odd function of $\nu$, but nonzero rotation introduces asymmetry, shifting the zero of $\tau_1$ away from $\nu=0$.

Fig.~1 shows the typical behavior of the Frenet=Serret scalars plotted versus relative velocity both on the equatorial plane $\theta=\pi/2$ and off ($\theta=\pi/4$) for an $a/m=1/2$ Kerr spacetime at radii approaching the black hole for the three radial intervals where two, one and then no timelike circular geodesics exist in the equatorial plane.
Figs.~1$a$) and 1$b$) show a kink in the curvature $\kappa$ in the equatorial plane at its zeros (geodesics), with a resultant sign change jump in $\tau_1$.
By allowing the Frenet-Serret vector $e_1$ to remain equal to $e_{\hat r}$ in the equatorial plane, while allowing $\kappa$ to assume negative values there as done in Figs.~5 and 9 of [\citen{idcf2}], both graphs become diffentiable at the location of the geodesics where the Frenet-Serret frame is only defined by continuity with the nearby accelerated orbits. This sign change for the torsion makes its new graph coincide with its absolute value $||\omega_{\rm(FS)}||$ (dotted curve) in the current plot.

In Fig.~1$a$) as one moves up off the equatorial plane at $\theta=\pi/2$, the three critical points of the curvature $\kappa$, all local extrema representing the two geodesics and the extremely accelerated orbit on that plane, all evolve, eventually resulting in only one critical point below a certain angle $\theta$ representing the single zero of the first torsion. Off the equatorial plane where it is identically zero, $\tau_2$ is a monotonic function of $\nu$ which has exactly one zero at the extreme value of the polar angle $\chi$ parametrizing the acceleration.

The fact that for sufficiently large positive $\nu$, $\tau_2$ is negative, and positive for sufficiently negative $\nu$ shows that $\chi$ is respectively increasing and decreasing in those limits, so the side of the hyperbola closer to the ``horizontal" dashed line in Fig.~2$a)$ of [\citen{circfs}] is the sufficiently positive $\nu$ side of the hyperbola, while the sufficiently negative side is pushed farther away from the horizontal, both of which continue rotating away from that horizontal direction as one approaches the horizon in Figs.~2$b)$ and 2$c)$. The true orientation of these hyperbolas is more apparent in the limiting Schwarzschild Fig.~5 of that article, where the chief affect of the gravitomagnetic field in the Kerr spacetime is to split the half-ray of gravitoelectric acceleration vectors into the two halves of the hyperbola branch, with the corotating half closer to the horizontal sufficiently far away from the horizon; the hyperbola rotates up and away from the vertical direction and outwards towards the radial direction as the horizon is approached. This is simply because the curvature vector $\vec k_{\rm(lie)}$ is well-behaved while the ZAMO acceleration $\vec a(n)$ ($\vec A$ in that figure) blows up as the horizon approaches and so dominates their linear combinations, and the latter acceleration field is approximately in the outward radial direction \cite{circfs} to keep a test particle from falling towards the ``center of attraction" when ``at rest with respect to the ZAMOs." This is the reason the electric part of the connection must dominate the magnetic part as one approaches the horizon in the Kerr spacetime: modulo common gamma factors the centripetal effects associated with the dominant term in the magnetic part are bounded relative to the growing ZAMO acceleration which dominates the electric part (see Eq.~(\ref{eq:EB(n)})).

Fig.~2 shows instead the two invariants at selected radii on and off the equatorial plane.
The graphs of $I_2$ in the equatorial plane show how the boost zone where it is positive expands as one approaches the horizon, first going superluminal for counterrotating orbits and then later for corotating orbits. $\tau_2$ and hence $I_1$ is identically zero in the equatorial plane, but off the equatorial plane, $I_1$ is an increasing function which grows faster as one approaches the horizon, always passing through zero at a very small negative relative velocity not visibly different from zero at this scale where $\tau_2=0$.

\typeout{Figure 2}
\begin{figure}[!t]
\typeout{*** EPS figure 2-1, 2-2}
\centerline{
\hbox{
\parbox{6cm}{\hbox to 6cm{\hfil $I_2$: $\theta=\pi/2$\hfil}
\epsfysize=6cm\epsfbox{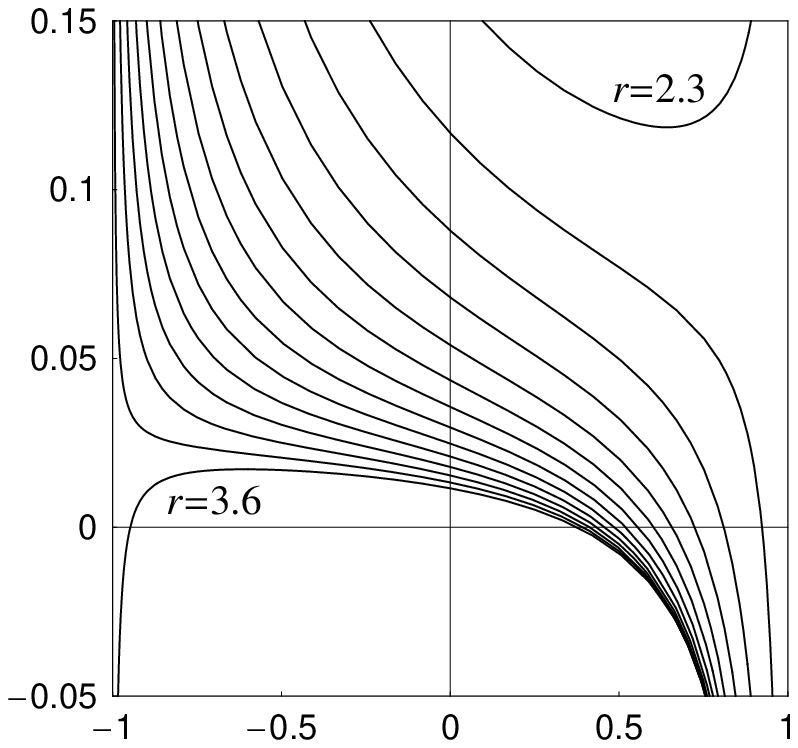}}
\hfill
\parbox{6cm}{\hbox to 6cm{\hfil $I_1$: $\theta=\pi/4$\hfil}
\epsfysize=6cm\epsfbox{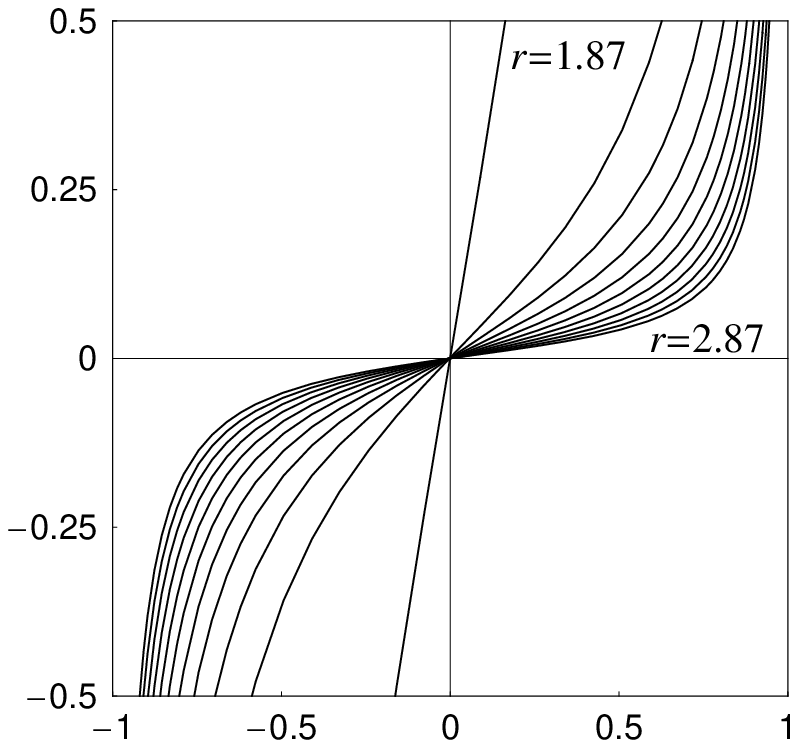}}
}
}
\caption{
The invariant $I_2$ plotted versus relative velocity at $\theta=\pi/2$ where $I_1=0$
for an $a/m=1/2$ Kerr spacetime at selected radii (the situation at $\theta=\pi/4$ is not much different qualitatively) and the invariant $I_1$
at $\theta=\pi/4$.
}
\label{fig:2}
\end{figure}

\section{The general solution of the transport equation along future-pointing timelike circular orbits}

As shown in section 2, the solution of the parallel transport
equation is (\ref{solution}), where $A$ is given by (\ref{matra})
and $\lambda=\tau$ is the proper time parametrization of the
orbit when $U$ is timelike.
$A$ is a tensor field only defined along the orbit
and the components of $A$ can be referred to any observer-adapted
frame once the observer family itself is chosen all along
the orbit. Splitting $A$ into its electric part associated with
a boost in an observer-adapted frame and a magnetic part
associated with a rotation
\beq A^\flat = u^\flat \wedge
\mathcal{E}(u) + \dualp{u} \mathcal{B}(u)
=A_{\rm B}(u)^\flat +A_{\rm R}(u)^\flat\ ,
\eeq
can be done with respect to any stationary unit timelike future-pointing vector field $u$ representing the 4-velocity of a stationary observer family defined along the orbit, but the separation is observer-dependent.

In the general Synge nonsingular case, $A$
generates a simultaneous boost and spatial rotation acting in orthogonal
2-planes, but the frame is not adapted to these 2-planes so the
boost and rotation are mixed together in their effects on the frame vectors.
The two parts $A_{\rm B}(u)$ and $A_{\rm R}(u)$ do not in general commute, so
$e^{A\, \tau}\neq e^{A_{\rm B}(u)\,\tau}\,e^{A_{\rm R}(u)\,\tau}$.
However, one can always find one observer for which these two parts do commute,
which in turn corresponds to parallel electric and magnetic fields
$\mathcal{E}(u)$ and $\mathcal{B}(u)$.
This has been explained in detail in [\citen{mtw}]
(exercise 20.6, p.~480) for a nonnull electromagnetic field where a new
observer with relative velocity along the direction of the Poynting
vector can be chosen to align the electric and magnetic
fields. We use that same procedure here.

For simplicity assume $U$ is a future-pointing timelike unit vector.
Start with
$A^\flat = U^\flat \wedge\mathcal{E}(U) + \dualp{U} \mathcal{B}(U)$ as seen by $U$ itself. Pick a new circularly rotating observer with 4-velocity
$\mathcal{X}(U)$ traveling in in the direction of $\mathcal{E}(U)\times_U
\mathcal{B}(U)$ with signed relative velocity $\tanh\tilde\alpha$
along the $\phi$ direction within the local rest space of $U$
chosen so that
\beq
|\tanh\,2\,\tilde\alpha|
= \frac{2\,||\mathcal{E}(U)\times_n \mathcal{B}(U)||}
       {\mathcal{E}(U)^2+\mathcal{B}(U)^2}\ .
\eeq
As long as the right hand side does not equal 1 (the Synge singular case for $A$), this has a real solution.
With a convenient choice of sign, direct evaluation using Eq.~(\ref{eq:EBtauomega}) gives
\beq
\label{alpha}
\tanh\,2\,\tilde\alpha
= \frac{2\,\kappa\,\tau_1}{\kappa^2 +||\omega_{\rm(FS)}||^2 }\equiv T\ .
\eeq
while identities then lead to
\begin{eqnarray}
\tanh \,\tilde\alpha = \frac{1-\sqrt{1-T^2}}{T}\equiv t\ ,\
\sinh\,\tilde\alpha = \frac{t}{\sqrt{1-t^2}} \ ,\
\cosh\,\tilde\alpha = \frac{1}{\sqrt{1-t^2}}\ .
\end{eqnarray}

Note that the sign of $\tilde\alpha$ is the same as the sign of $\tau_1$ as long as $\kappa\geq0$ as assumed here, while $\tilde\alpha$ itself reduces to 0 in the limit $\tau_1\to0$ or $\kappa\to0$ and the Frenet-Serret frame is itself adapted to the parallel transport boost and rotation.
The first case of vanishing first torsion corresponds to the extremely accelerated orbits; 
when in addition the second torsion goes to zero as it does identically in the equatorial plane of the Kerr spacetime, this becomes a pure boost.
The second case of vanishing curvature corresponds to the geodesics in the equatorial plane, which instead leads to a pure rotation in the local rest space of the geodesic.
When $\tilde\alpha\neq0$, its sign indicates whether one must slow down ($-$) or speed up ($+$) in the positive $\phi$-direction relative to $U$ to achieve the canonical form for the connection matrix $A^\alpha{}_\beta$.

Fig.~3 illustrates the behavior of the relative velocity
$\tanh\tilde\alpha$. Far enough from the horizon where there are
two geodesics in the equatorial plane, $\tilde\alpha$ is zero at
the two geodesics and at the extremely accelerated orbit in
between them, but as one moves in, first the extremely accelerated
orbit and the counterrotating geodesic go null together, leaving
one zero, and then finally the corotating geodesic goes null (not
yet reached in this sequence). Similarly the values $\pm1$ occur
when the Synge singular orbits are reached at the endpoints of the
boost zone, which grows outwards from $\nu=0$ as one moves towards
the horizon, exiting first at $\nu=-1$ from the timelike interval.
Off the equatorial plane, there is only one zero at the extremely
accelerated orbit. For velocities faster than this, one must slow
down, and for velocities less than this, speed up, to boost $U$ by
this relative velocity.

\typeout{Figure 3}
\begin{figure}[t]
\typeout{*** EPS figure 3-1, 3-2}
\centerline{
\hbox{
\parbox{6cm}{\hbox to 6cm{\hfil $\theta=\pi/2$\hfil}
\epsfysize=6cm\epsfbox{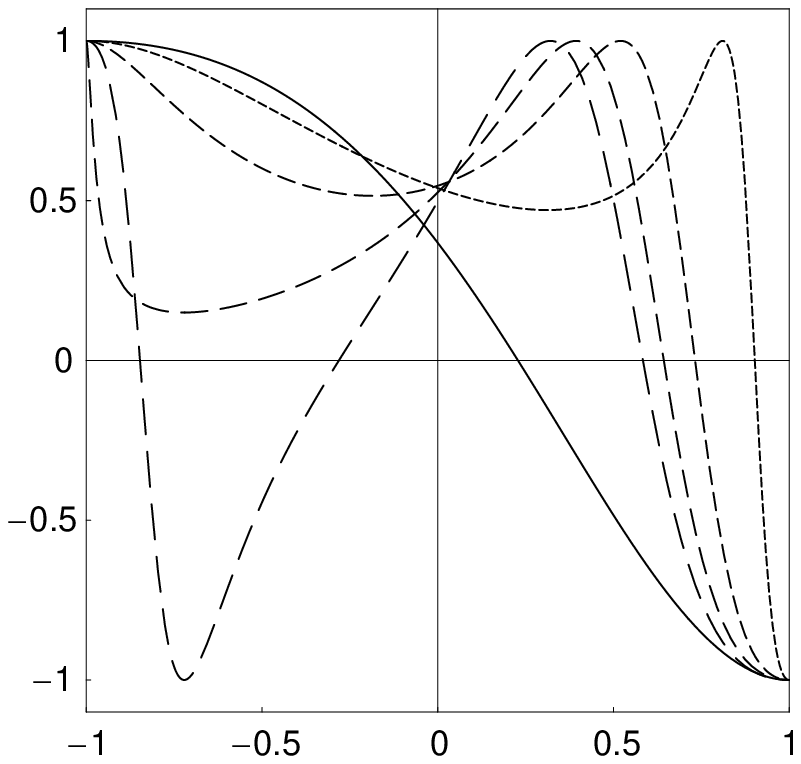}}
\hfill
\parbox{6cm}{\hbox to 6cm{\hfil $\theta=\pi/4$\hfil}
\epsfysize=6cm\epsfbox{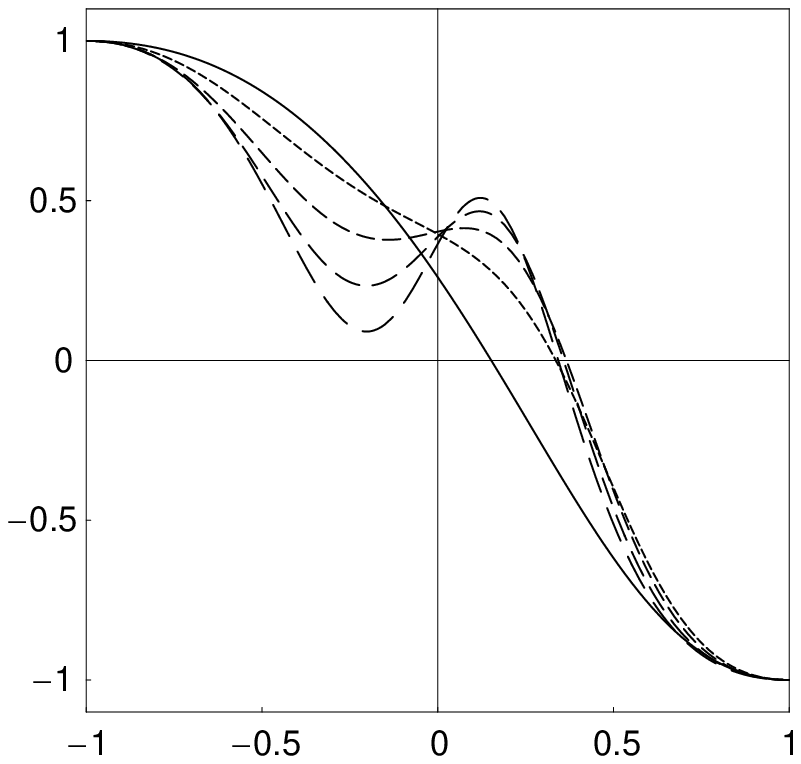}}
}
}
\caption{The boost velocity ${\rm tanh}\,\tilde\alpha$ from $U$ to the parallel transport adapted frame plotted versus $\nu$, shown for
 $r=2,\, 2.5,\, 3,\, 3.4,\, 4$ (from solid to increasingly longer dashed lines) in the $a/m=1/2$ Kerr spacetime.
}
\label{fig:3}
\end{figure}

Boosting the Frenet-Serret frame vectors $\{U,\,e_2\}$ to
$\{E_0,\,E_2\}$ by the relative observer boost from $U$ to
$\mathcal{X}(U)$ and defining the second pair of frame vectors
$E_1$ and $E_3$ as proportional to their respective covariant
derivatives gives the form
\begin{eqnarray}
E_0
&=& \mathcal{X}(U)
= \cosh\tilde\alpha\,U+\sinh\tilde\alpha\,e_2
\ ,\nonumber\\
E_2
&=& \bar\mathcal{X}(U)
= \sinh\tilde\alpha\,U+\cosh\tilde\alpha\,e_2
\ ,\nonumber\\
E_1
&=& \sigma_{\rm B}^{-1}\,
[ (\kappa\,\cosh\tilde\alpha - \tau_1\,\sinh\tilde\alpha)\,e_1
           + \tau_2\,\sinh\tilde\alpha\,e_3]
\ ,\nonumber\\
E_3
&=& \sigma_{\rm R}^{-1}\,
[ (\kappa\,\sinh\tilde\alpha - \tau_1\,\cosh\tilde\alpha)\,e_1
       + \tau_2\,\cosh\tilde\alpha\,e_3]\ ,
\end{eqnarray}
(reducing to the identity transformation when $\tilde\alpha=0$)
which leads to
\beq\label{eq:Asigmas}
A^\flat = -\sigma_{\rm B}\,E_0^\flat \wedge E_1^\flat
           + \sigma_{\rm R}\,E_2^\flat\wedge E_3^\flat\ .
\eeq Requiring orthogonality of $E_1$ and $E_3$ leads to the
condition (\ref{alpha}), while the invariants (recall the
relations (\ref{eq:syngekt}))
\beq
I_1 = -\sigma_{\rm B}\,
\sigma_{\rm R} = -\kappa\,\tau_2\ ,\ I_2 = \sigma_{\rm B}^2 -
\sigma_{\rm R}^2 = \kappa^2-(\tau_1^2+\tau_2^2)
\eeq
may then be
used to determine the nonnegative boost parameter $\sigma_{\rm
B}\geq0$ and the rotation parameter $\sigma_{\rm R}$ in terms of
the curvature and torsions \beq\label{eq:sigmas} \sigma_{\rm
B}{}^2 = [(I_2{}^2+I_1{}^2)^{1/2} + I_2]/2\ ,\ \sigma_{\rm R}{}^2
= [(I_2{}^2+I_1{}^2)^{1/2} - I_2]/2\ , \eeq equivalent to
normalizing $E_1$ and $E_2$. Note that $E_2=\bar\mathcal{X}(U)$ is
just the unit vector in the direction of increasing $\phi$ in the
local rest space of $\mathcal{X}(U)$ and the sign of $\sigma_R$
must agree with the sign of $\tau_2$.

In the Synge semi-singular case $I_1=-\kappa\,\tau_2=0$ but
$I_2\neq0$, then parallel transport reduces to a pure boost
($I_2>0 \rightarrow \sigma_R=0,\, \sigma_B=|I_2|^{1/2}$) under
which $\bar\mathcal{X}(U)$ is invariant or a pure rotation ($I_2<0
\rightarrow \sigma_B=0,\,\sigma_R=|I_2|^{1/2}$) under which
$\mathcal{X}(U)$ is invariant. This is true in the equatorial
plane of the Kerr spacetime where $\tau_2=0$ identically and this
invariant unit vector $\bar\mathcal{X}(U)$ or $\mathcal{X}(U)$
respectively defines the equatorial plane map $U\to\mathcal{Z}(U)$
which picks out the direction of the covariant constant vector in
the velocity plane along the orbit \cite{bjm2}. For geodesics in
the equatorial plane, this is a pure rotation in the local rest
space of $U$ as remarked above. Figs.~4 and 5 show the typical
behavior of $\sigma_B$ and $\sigma_R$.

\typeout{Figure 4}
\begin{figure}[t]
\typeout{*** EPS figures 4-1, 4-2}
\centerline{
\hbox{
\parbox{6cm}{\hbox to 6cm{\hfil $\theta=\pi/2$\hfil}
\epsfysize=6cm\epsfbox{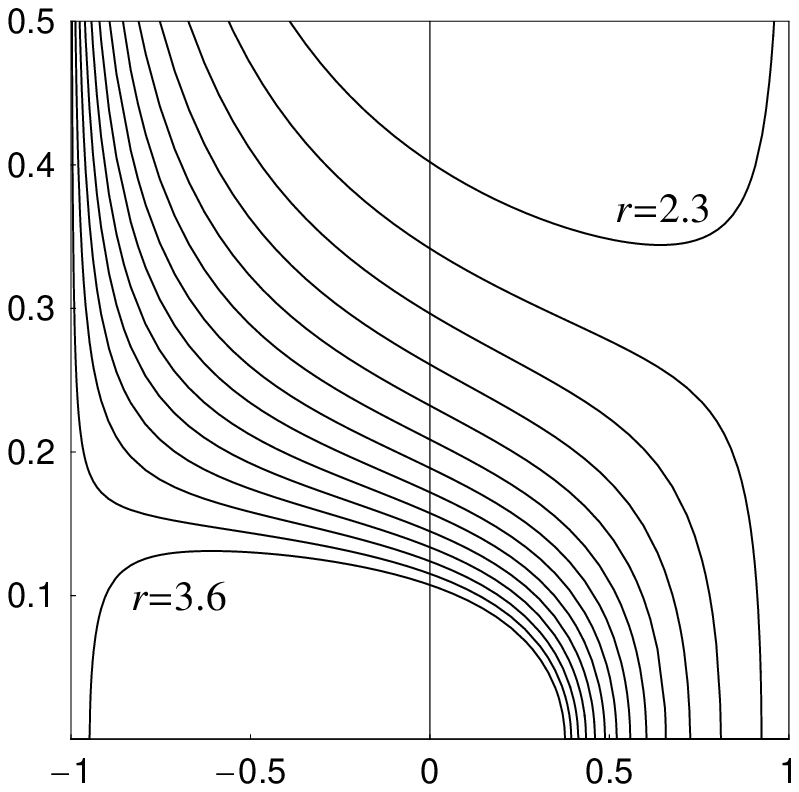}}
\hfill
\parbox{6cm}{\hbox to 6cm{\hfil $\theta=\pi/4$\hfil}
\epsfysize=6cm\epsfbox{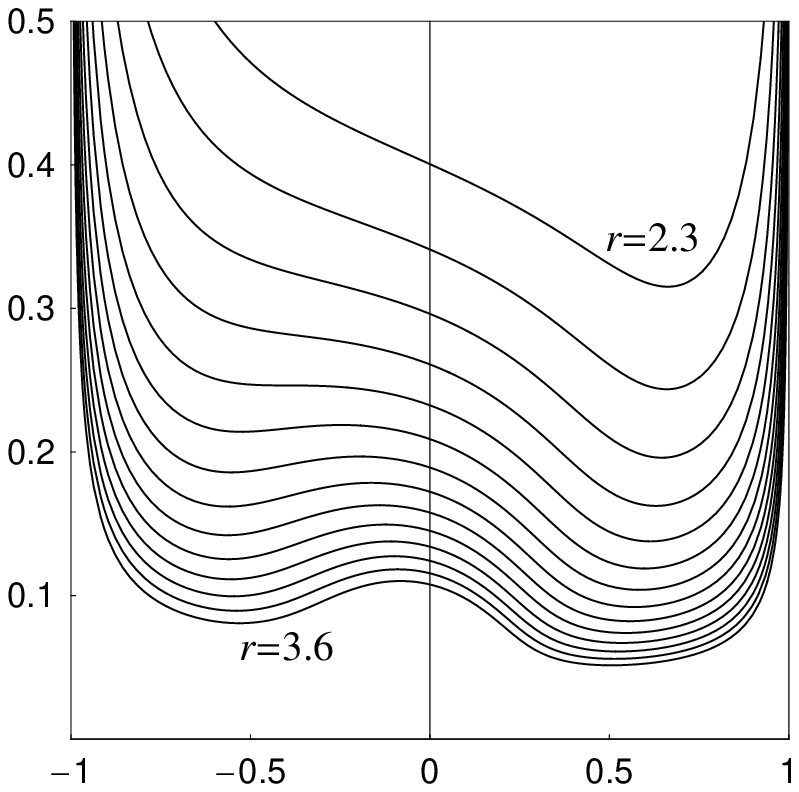}}
}
}
\caption{$\sigma_B$ versus $\nu$ for the
$a/m=1/2$ Kerr spacetime.
}
\label{fig:4}
\end{figure}

\typeout{Figure 5}
\begin{figure}[t]
\typeout{*** EPS figures 5-1, 5-2}
\centerline{
\hbox{
\parbox{6cm}{\hbox to 6cm{\hfil $\theta=\pi/2$\hfil}
\epsfysize=6cm\epsfbox{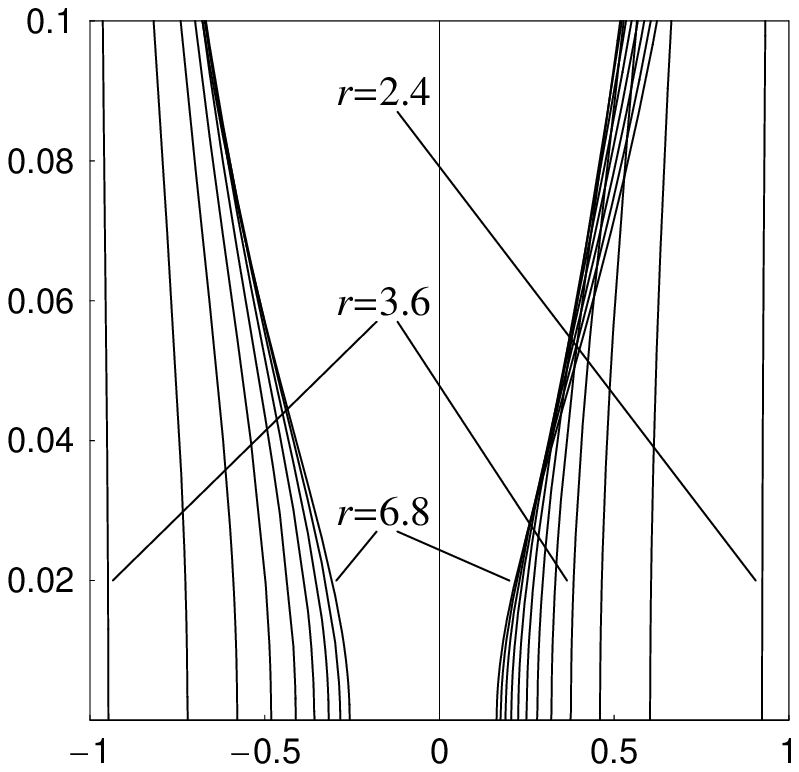}}
\hspace{6mm}
\parbox{6cm}{\hbox to 6cm{\hfil $\theta=\pi/4$\hfil}
\epsfysize=6cm\epsfbox{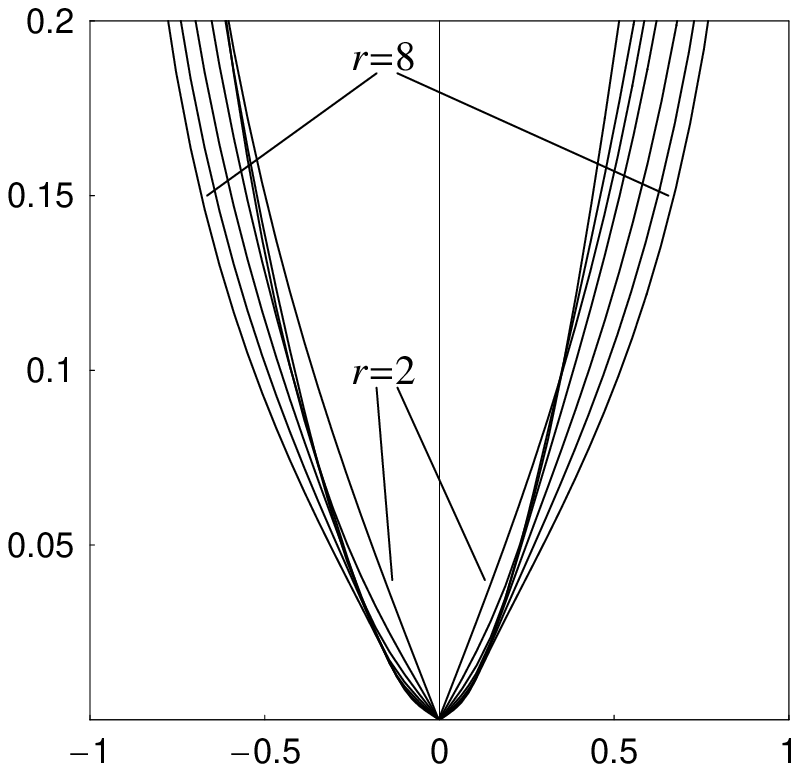}}
}
}
\caption{$\sigma_R$ versus $\nu$ for the
$a/m=1/2$ Kerr spacetime.}
\label{fig:5}
\end{figure}

To express the boosted frame in terms of the original symmetry
adapted ZAMO frame, one can follow the same approach, starting
with a boost \beq
 E_0 = \mathcal{X}(U)
     = \cosh\hat\alpha\, n +\sinh\hat\alpha\,e_{\hat\phi}
\ ,\
 E_2 = \bar\mathcal{X}(U)
     = \sinh\hat\alpha\, n +\cosh\hat\alpha\,e_{\hat\phi}
\eeq
and using its inverse to re-express
\beq
  A^\flat = n^\flat \wedge \mathcal{E}(n) + \omega^{\hat\phi}\wedge  \mathcal{K}\ ,\
\mathcal{K} = \gamma\,\nu\, k_{\rm(fw)} \eeq in the form \beq
  A^\flat = \mathcal{X}(U)^\flat \wedge \mathcal{E}(\mathcal{X}(U))
           + \bar\mathcal{X}(U)^\flat \wedge \mathcal{K}(\mathcal{X}(U))\ ,
\eeq
where
\beq
 \mathcal{E}(\mathcal{X}(U))
       = \cosh\hat\alpha\, \mathcal{E}(n) -\sinh\hat\alpha\, \mathcal{K}\ ,\
 \mathcal{K}(\mathcal{X}(U))
       = -\sinh\hat\alpha\, \mathcal{E}(n) +\cosh\hat\alpha\, \mathcal{K}\ .
\eeq
Then the orthogonality condition (alignment of electric and magnetic vectors)
\beq
0 =  \mathcal{E}(\mathcal{X}(U)) \cdot  \mathcal{K}(\mathcal{X}(U))
  = \bar\mathcal{X}(U) \cdot
    [\mathcal{E}(\mathcal{X}(U)) \times_{\mathcal{X}(U)} \mathcal{B}(\mathcal{X}(U))]
\eeq
yields the corresponding rapidity condition
\beq
  \tanh2\hat\alpha = \frac{\mathcal{E}(n)\cdot
     \mathcal{K}}{||\mathcal{E}(n)||^2 + ||\mathcal{K}||^2}
    = \frac{e_{\hat\phi} \cdot[\mathcal{E}(n)\times_n \mathcal{B}(n)]}
         {||\mathcal{E}(n)||^2 + ||\mathcal{B}(n)||^2}\ .
\eeq
from which one can determine the ZAMO relative velocity
$\nu_{\mathcal{X}}=\tanh\hat\alpha = [1-(1-\tanh^2 2\hat\alpha)^{1/2}]/\tanh 2\hat\alpha$ of $\mathcal{X}(U)$
in terms of $\vecA$, $\vec k_{\rm(lie)}$, and $\vec\theta_{\hat\phi}$ using (\ref{eq:EB(n)}).

Now that the boost velocity is determined with respect to either the ZAMOs or $U$ itself, consider the component matrix $(A^\alpha{}_\beta)$ in the new boosted frame $E_\alpha$, which
has the canonical block-diagonal form
\beq (A^\alpha{}_\beta)
= (A_{\rm B}{}^\alpha{}_\beta) + (A_{\rm R}{}^\alpha{}_\beta)
= \left(
\begin{array}{cc} \left[
  \begin{array}{cc}
   0 & \sigma_{\rm B} \\[1mm]
   \sigma_{\rm B} & 0
  \end{array} \right]\hspace{3mm}
& \mbox{\Large0}_2 \vspace{3mm}\cr \mbox{\Large 0}_2
& \mbox{\Large0}_2\end{array}\right)
+
\left(
\begin{array}{cc}
\mbox{\Large 0}_2 & \mbox{\Large 0}_2 \vspace{3mm}\cr
\mbox{\Large 0}_2
& \hspace{3mm}\left[
  \begin{array}{cc}
   0 & \sigma_{\rm R} \\[1mm]
   -\sigma_{\rm R} & 0
   \end{array} \right]
\end{array}\right)
\eeq where $A_{\rm B}$ has eigenvalues $\lambda=0,\,0,\,\pm
\sigma_{\rm B}$ and $A_{\rm R}$ has eigenvalues
$\lambda=0,\,0,\,\pm i \,\sigma_{\rm R}$. This translates the
action of parallel transport along the orbit for a generic vector
$X$ into two commuting actions of a family of boosts and rotations
when expressed in this frame \beq \dot{X}(\tau) = A\, X(\tau)\ ,
\quad \rightarrow \quad X(\tau) = e^{A_{\rm B}\,\tau}\,e^{A_{\rm
R}\,\tau}\,X(0)\ , \eeq so that the frame components of a parallel
transported vector $X(\tau)$ in this new frame are only coupled
within two mutually orthogonal subspaces \beq \pmatrix{X^0(\tau)
\cr X^1(\tau)} = \pmatrix{\cosh\sigma_{\rm B}\,\tau &
\sinh\sigma_{\rm B}\,\tau \cr
 \sinh\sigma_{\rm B}\,\tau & \cosh\sigma_{\rm B}\,\tau}\,
\pmatrix{ X^0(0) \cr X^1(0)}
\eeq
\beq
\pmatrix{X^2(\tau) \cr X^3(\tau)}
=
\pmatrix{\cos\sigma_{\rm R}\,\tau & \sin\sigma_{\rm R}\,\tau \cr
  -\sin\sigma_{\rm R}\,\tau & \cos\sigma_{\rm R}\,\tau}\,
\pmatrix{X^2(0) \cr X^3(0)}
\eeq
and evolution of the initial data in the general case is governed
respectively by a boost or a rotation in these subspaces.
In the general nonsingular case, both must be present.

When $A$ is singular in the Synge sense, corresponding to
vanishing values of the quantities $\sigma_{\rm B}$ and
$\sigma_{\rm R}$ defined by Eqs.~(\ref{eq:sigmas}) the above
derivation and the representation (\ref{eq:Asigmas}) do not hold
and must be redone. Circular orbits for which $A$ is singular but
nonzero cannot be geodesics but must have nonzero $\kappa=\pm
\tau_1$ and $\tau_2=0$. Then the mixed component matrix of $A$
with respect to $\{u,e_1,e_2,e_3\}$ is \beq A^\flat = \kappa\, e_1
\wedge [U\pm e_2], \qquad (A^\alpha{}_\beta) = \pmatrix{ 0 &
\kappa & 0 & 0 \cr \kappa & 0 & \pm\kappa & 0 \cr 0 & \pm\kappa &
0 & 0 \cr 0 & 0 & 0 & 0 \cr }\ , \eeq which generates a null
rotation \beq e^{A\tau}= 1+A\,\tau +\frac12\, A^2\, \tau^2\ \eeq
in the plane orthogonal to $e_3$.

In the Kerr spacetime, this Synge singular case only occurs in the
equatorial plane as the transition between the parallel transport
boost and rotation velocity zones for each circular orbit
\cite{bjm2}. That $A^\flat$ is a decomposable 2-form for the Synge
singular case is true for the entire Synge semi-singular case
which characterizes the Kerr equatorial plane where the vanishing
of $\tau_2$ makes $\sigma_B\, E_1$ and $\sigma_R\, E_3$ both
proportional to $e_1$ (so only one term survives in
Eq.~(\ref{eq:Asigmas}), proportional to $e_1\wedge E_0$ or
$e_1\wedge E_2$), which then factors out of $A^\sharp$ as one also
sees directly from setting $\tau_2=0$ in Eq.~(\ref{eq:EBtauomega})
\beq
  A^\sharp = -\kappa\, U\wedge e_1 +\tau_1\, e_1\wedge e_2
           = e_1 \wedge (\kappa \,U + \tau_1\, e_2)\ .
\eeq
When the vector in parentheses is nonnull, it is proportional to $E_0=\mathcal{X}(U)$ if timelike ($I_1>0$) corresponding to a parallel transport boost or $E_2=\bar\mathcal{X}(U)$ if spacelike ($I_2<0$) corresponding to a parallel transport rotation, and taking the parametrization relationship (\ref{eq:phi2tau}) into account, while choosing instead $e_1=e_{\hat r}$ (allowing $\kappa$ to change sign),
one can translate this back to the map $\mathcal{Z}$ determining the invariant parallel transport direction using (A.7) of [\citen{bjm2}].

\section{Boost and rotation domination zones}

The Synge class of the matrix $A$ associated with the invariants
$I_1$ and $I_2$ is a function of the two nonignorable coordinates
$r$ and $\theta$. Either $I_2\neq0$ (nonsingular case) or $I_1=0$
but $I_2\neq0$ (semi-singular case) or   $I_1=0=I_2$ (singular
case), each case corresponding to a different class of Lorentz
transformations governing parallel transport.

The sign of $I_2$ indicates the relative magnitudes of the boost
and rotation rates along a circular orbit in the nonsingular case
(boost-dominated if $I_2>0$, rotation-dominated if $I_2<0$) and
whether a pure boost ($I_2>0$) or pure rotation ($I_2<0$) occurs
in the semi-singular case. Such boost-dominated and
rotation-dominated zones occur in each family of circular orbits
at a given radius and angular location as a function of the
relative velocity, but as one approaches the vicinity of the
horizon in a Kerr spacetime, for timelike circular orbits the
rotation-dominated zones disappear as the attraction to the black
hole (acceleration) overcomes any relative centripetal
acceleration effects (Frenet-Serret angular velocity) due to the
limits on the relative velocity $|\nu|<1$.

For a given radial and angular position of a circular orbit, the
condition $I_2=0$ determines two relative velocities which delimit
an interval $\nu_{(PT-)} < 0 < \nu_{(PT+)}$ around the ZAMO
angular velocity $\nu=0$ within which parallel transport
boost-dominance occurs ($I_2>0$) \beq
 |\kappa| > ||\omega_{\rm FS}||
\leftrightarrow
 ||\mathcal{E}(n)|| > ||\mathcal{B}(n)||
\leftrightarrow
 || F^{(G)}_{\rm(fw)} || > \gamma\,\nu ||k_{\rm(fw)}||\ .
\eeq
and the Fermi-Walker relative gravitational force on the circular orbit exceeds the Fermi-Walker relative angular velocity.
Using (\ref{eq:EB(n)}) one can express this inequality as a quadratic inequality in $\nu$
\begin{eqnarray}\label{eq:quad}
I_2&=& (||\theta_{\hat\phi}||^2 -||k_{\rm(lie)}||^2)\, \nu^2 -2\,
\theta_{\hat\phi}\cdot(g+k_{\rm(lie)})\, \nu +||g||^2 -
||\theta_{\hat\phi}||^2
\nonumber\\
&=& -\mathcal{A}\, \nu^2 -2\,\mathcal{B} \,\nu +\mathcal{C} > 0
\end{eqnarray}
whose corresponding equality has the solutions $\nu_{\rm(PT-)}
\leq \nu_{\rm(PT+)}$ given by the minimum and maximum values of
the usual quadratic formula for the roots \beq
  \nu_{\rm(PT\pm)}
   = \frac{-\mathcal{B}\pm [\mathcal{B}^2
   +\mathcal{A}\,\mathcal{C})]^{1/2}}{\mathcal{A}}\ .
\eeq
This leads to the static limit
\beq
  \theta_{\hat\phi} \to0: \quad
    \nu_{\rm(PT\pm)}
    \to \frac{||\vec g||}{||k_{\rm(lie)}||}
       \left(\pm1-\frac{\theta_{\hat\phi}\cdot(g+k_{\rm(lie)})}
                       {||g||\, ||k_{\rm(lie)}||}\right)\ ,
\eeq
with $\mathcal{B}=\theta_{\hat\phi}\cdot(g+k_{\rm(lie)})>0$ for a corotating ($a>0$) Kerr spacetime \cite{circfs} and $\mathcal{A}=||k_{\rm(lie)}||^2-||\theta_{\hat\phi}||^2  >0$ holds everywhere except very close to the horizon.

When the relative velocity moves out of the interval
$[\nu_{\rm(PT-)},\, \nu_{\rm(PT+)}]$, the circular orbit has
sufficient speed for the relative Fermi-Walker angular velocity to
exceed the Fermi-Walker spatial gravitational force. Inside this
interval the Fermi-Walker spatial gravitational force dominates
the relative angular velocity. At these endpoint values, the
circular orbit parallel transports one of the two circular null
tangent vectors, as shown by Eq.~(\ref{eq:E=B}); in the equatorial
plane, $A$ is then singular generating a null rotation as
described in [\citen{bjm2}].

The condition $\nu_{\rm(PT\pm)} =\pm1 $, after clearing fractions,
isolating the radical and squaring both sides of the resulting
equation, becomes $\mathcal{A}\,(\mathcal{A} \pm 2\,\mathcal{B}
-\mathcal{C})=0$. Setting the second factor to zero yields 
\beq
      (\A - k_{\rm(lie)})\cdot
         (\A + k_{\rm(lie)} \pm 2\,\theta_{\hat\phi}) =0
 \leftrightarrow
      \vec a_0 \cdot \vec a_\pm =0
\eeq
which shows that these limiting null orbits occur when the position vector $\vec a_0=-\vec g_+$ of the center of the acceleration hyperbola is perpendicular to one of the two asymptote directions $\vec a_\pm$ corresponding to null orbits, or if $\vec a_0$ vanishes, or if one of those null circular orbit acceleration vectors vanishes. The latter corresponds to corotating and counterrotating null circular geodesics, which occur in the equatorial plane of the Kerr spacetime.

The limits $\nu_{\rm(PT\pm)} \to \pm\infty$ correspond to
$\bar\nu_{\rm(PT\pm)}=1/\nu_{\rm(PT\pm)} \to 0^{\pm}$ which
describe the spacelike closed $\phi$-coordinate circles. These
occur where the leading coefficient $-\mathcal{A} =
||\theta_{\hat\phi}||^2 -||k_{\rm(lie)}||^2$ goes to zero in the
above quadratic relationship (\ref{eq:quad}) for the root where no
cancellation occurs in the numerator in this limit. But for null
relative velocities $\nu=\pm1$, one has
$k_{\rm(fw)}=k_{\rm(lie)}+(\pm1)\,\theta_{\hat\phi}$, so this will
occur at locations $(r,\,\theta)$ where the Fermi-Walker curvature
for one of the two null orbits vanishes, corresponding to a
``Fermi-Walker relatively straight" curve, where the Fermi-Walker
relative centripetal acceleration goes to zero.

\begin{figure}[t]
\typeout{Figure 6}
\centerline{\epsfxsize=\textwidth
\epsfbox{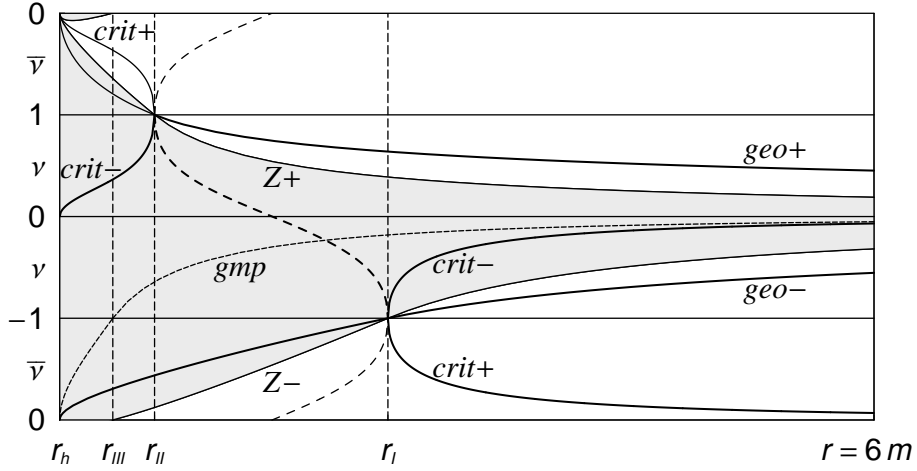}
} \caption{ The boost- (grey shading) and rotation-dominated
(white) relative velocity zones for the equatorial plane in the
$a/m=1/2$ Kerr spacetime for $-\infty < \nu < \infty$ using the
reciprocal velocity $\bar\nu=1/\nu$ to plot the spacelike relative
velocities $|\nu|>1$, with the top and bottom borders identified
to describe the same closed $\phi$ circles being traced out in
opposite directions. The velocity curves labeled by $Z_\pm
=\nu_{\rm(PT\pm)}$ separate these two zones. Vertical lines
indicate the radii at which the counterrotating ($r_{I}$) and
corotating ($r_{II}$) geodesics go null, the radius $r_{III}$
where the geodesic meeting point orbits go null and the horizon
$r_h$ where the null meeting point orbits (ZAMO world lines
outside the horizon, see [\citen{bjm01}]) go null. These
correspond respectively to the radii where $\nu_{\rm(PT-)}=-1$
($I$), $\nu_{\rm(PT+)}=1$ ($II$), $\nu_{\rm(PT-)}\to -\infty$
$(III$) and $\nu_{\rm(PT+)} \to \infty$. Also shown are the
relative velocity profiles of the geodesics (geo$\pm$), geodesic
meeting point curves (gmp), and the critical points of the
circular orbit acceleration where $\tau_1=0$  (crit$\pm$). The
subluminal critical point curve (crit$-$) corresponds to the
extremely accelerated observers. The dashed curve between the
radii $r_{I}$ and $r_{II}$ shows instead the critical points of
$\tau_1 =|\omega_{\rm(FS)}|$. } \label{fig:6}
\end{figure}

For a corotating Kerr spacetime, as one moves closer to the black
hole in the equatorial plane \cite{bjm2,fermi}, the interval
$[\nu_{\rm(PT-)},\, \nu_{\rm(PT+)}]$ expands until
$\nu_{\rm(PT-)}=-1$ at the counterrotating null circular geodesic
so that counterrotating rotation-dominated timelike orbits no
longer exist and then continues until finally $\nu_{\rm(PT+)}=1$
at the corotating null circular geodesic so that not even
corotating rotation-dominated timelike orbits exist beyond that
\cite{fermi}. In the equatorial plane these occur exactly at the
counterrotating ($r_{I}$) and corotating ($r_{II}$) null geodesic
orbits, while $\nu_{\rm(PT-)}\to-\infty$ at the geodesic meeting
point observer horizon ($r_{III}$) and $\nu_{\rm(PT+)}\to\infty$
at the black hole horizon (ZAMO horizon: $r_h$), corresponding to
transport around closed $\phi$ coordinate loops \cite{fermi}.
Fig.~\ref{fig:6} shows the boost-dominated zone (or just boost
zone since there is no simultaneous rotation) shaded in grey, with
the remaining rotation-dominated zone (or just rotation zone) in
white. This explains the domains of $\sigma_B$ (inside the boost
zone) and $\sigma_R$ (outside the boost zone) respectively in
Figs.~4$a)$ and 5$a)$, while off the equatorial plane both
quantities are defined for all velocities except $|\nu|\to1$ where
they both go infinite due to a gamma factor in their definition.

\begin{figure}[t]
\typeout{Figure 7}
\centerline{\epsfxsize=\textwidth
\epsfbox{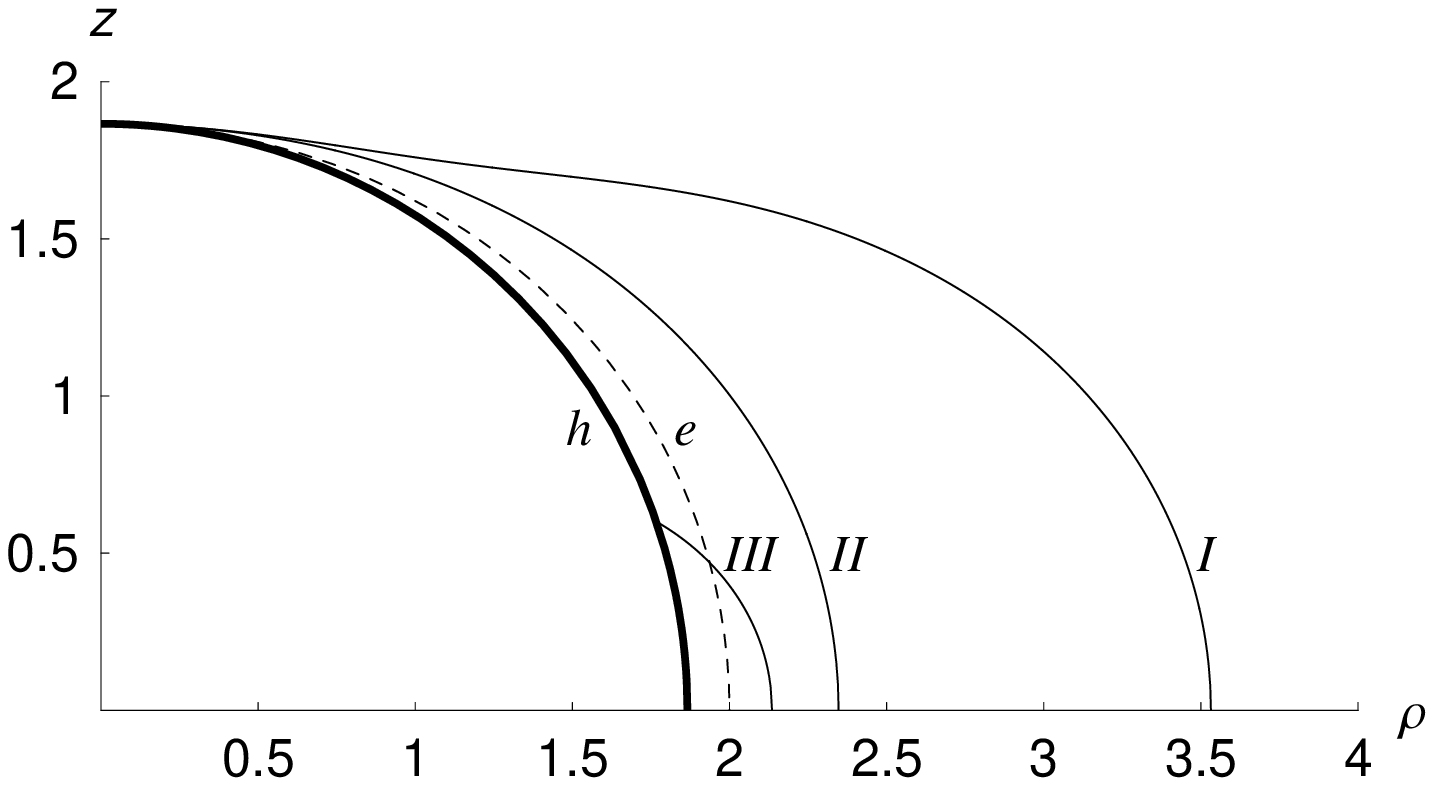}
} \caption{ A polar plot in the $r$-$\theta$ plane treated as
though they were polar coordinates in a flat 2-space with
horizontal axis $\rho=(r/m)\sin\theta$ and vertical axis
$z=(r/m)\cos\theta$, for an $a/m=1/2$ Kerr spacetime, showing the
event horizon (h, thick curve), where $\nu_{\rm(PT+)} \to \infty$,
the ergosphere boundary (e, dashed curve), the curve $I$ where
$\nu_{\rm(PT-)}=-1$ and the curve $II$ where $\nu_{\rm(PT+)}=1$ ,
and finally the curve $III$ intersecting the horizon at about
$19^o$ above the equatorial plane, along which $\nu_{\rm(PT-)}\to
-\infty$ from the outside, but $\infty$ from the inside (see
Fig.~\ref{fig:6}). } \label{fig:7}
\end{figure}

Fig.~\ref{fig:7} extends these boundary regions off the equatorial
plane to show the horizon, ergosphere boundary and the curves in
an $r$-$\theta$ section of the exterior Kerr spacetime where
$\nu_{\rm(PT\pm)}$ take the values $\pm 1$ and have the limits
$\pm\infty$. The first two conditions define two closed surfaces
of revolution surrounding the horizon and ergosphere boundary, the
II surface (corotating) inside the I surface (counterrotating),
which both meet the horizon and ergosphere at the symmetry axis
$\theta=0$ but extend farther out from them and each other as one
approaches the equatorial plane, which they intersect in circles
at the respective radii where the corotating null geodesics and
counterrotating null circular geodesics exist. The limit
$\nu_{\rm(PT-)}\to-\infty$ (approaching from the side away from
the origin, but $\infty$ when approaching from inside) occurs
along the smaller curve III in Fig.~\ref{fig:7} near the
equatorial plane up to about $\theta=71^o$ where it intersects the
horizon, where in turn the limit $\nu_{\rm(PT+)}\to\infty$ occurs.

In the static Schwarzschild limit, the surfaces corresponding to
the curves $I$ and $II$ in Fig.~\ref{fig:7} come together at the
radius $r=3\,m$ in the equatorial plane, while the horizon,
ergosphere boundary and the infinite limit surfaces collapse to
the sphere $r=2\,m$ (see Fig.~2 of \citen{bjm2}). Inside the
single surface $I = II$, no timelike timelike circular orbit
exists with rotation-domination. In the general Kerr spacetime,
inside of the outermost surface $I$, no timelike counterrotating
circular orbit exists with parallel transport rotation-domination,
while inside the next surface $II$, no timelike corotating or
counterrotating timelike circular orbit exists with
rotation-domination. Inside the next surface $III$, no
counterrotating circular orbit of any causality type exists with
rotation-domination, while at the horizon all circular orbits of
any kind are boost-dominated as the rotation-domination velocity
zone shrinks to the null set at the closed $\phi$-coordinate
circle being traversed in the corotating direction of increasing
$\phi$, exactly as in the equatorial plane in Fig.~\ref{fig:6}.

\begin{figure}[!t]
\typeout{Figure 8} \centerline{$a)$ $\theta=1.5$}
\centerline{\epsfxsize=\textwidth \epsfbox{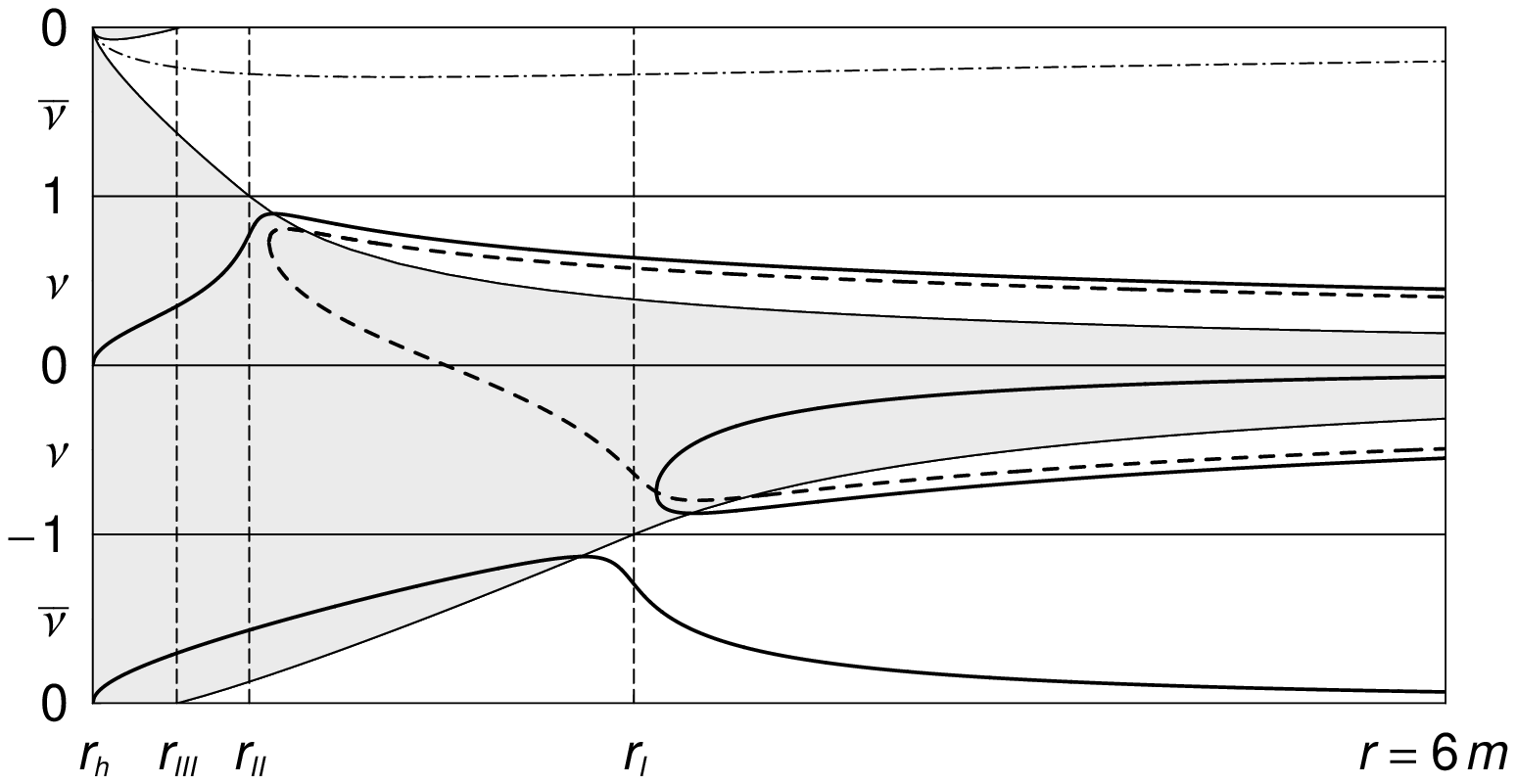}}
\centerline{$b)$ $\theta=\pi/3$} \centerline{\epsfxsize=\textwidth
\epsfbox{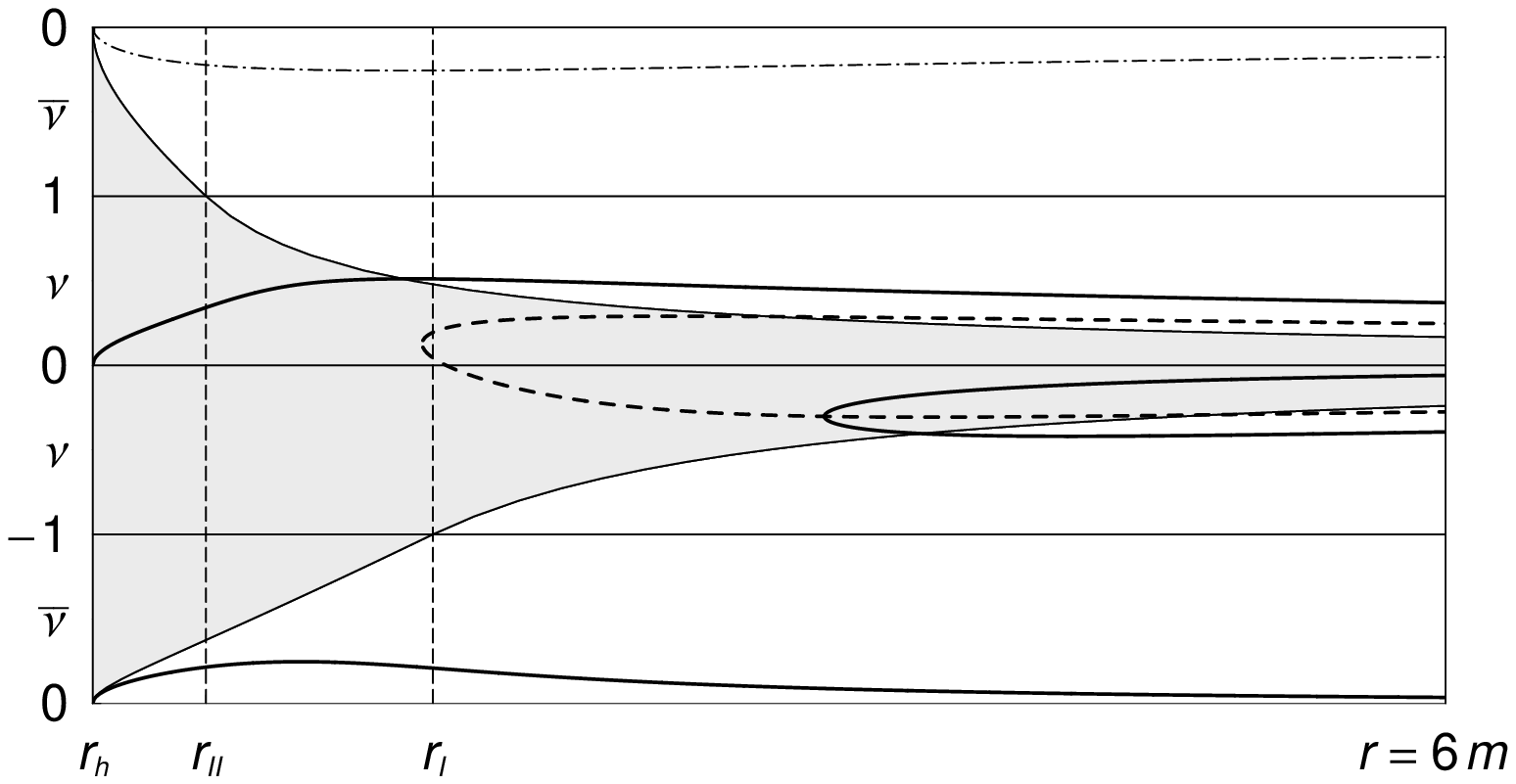}}
\caption{ The figure corresponding to
Fig.~\ref{fig:6} off the equatorial plane ($\theta=\pi/2\approx
1.57$) nearby at $\theta=1.5$ and farther away at $\theta=\pi/3$.
Vanishing $\tau_1$ occurs at the solid curves (not bounding the grey region),
while the dashed
curve indicates the zero of its $\nu$-derivative. In $a)$ one sees
that departing from the equatorial plane, the geo$-$ and
counterotating crit$-$ curves to the right of $r_{I}$ in the equatorial plane
Fig.~\ref{fig:6} deform and join together to produce the
subluminal counterrotating vanishing $\tau_1$ curve,
while the geo$+$ curve
and the corotating crit$-$ curve similarly produce the corotating
such subluminal curve; the counterrotating superluminal crit$+$ curve
joins together with the superluminal geo$-$ curve to produce the
the superluminal counterrotating vanishing $\tau_1$ curve.
The boost zone shrinks with decreasing angle. The dashed-dotted
uppermost curve in the white rotation-dominated zone corresponds
to vanishing $\tau_2$ at positive superluminal speeds (converging
with the boost-dominated zone boundary curves at the horizon at
$\bar\nu=0$), while the subluminal such curve occurs at very small
negative relative velocities not distinguishable from the
$\nu$-axis at this scale.} \label{fig:8}
\end{figure}

One can imagine how this latter figure changes as one decreases
$\theta$ from $\pi/2$ in Fig.~\ref{fig:6}. Fig.~\ref{fig:8}$a)$
shows how the equatorial plane velocity curves deform and join to
produce the vanishing $\tau_1$ velocity curves as the boost zone
squeezes down towards $\nu=0$ and its marking radii move in
towards the horizon as $\theta$ decreases. This continues in $b)$,
where the radius $r_{III}$ has already disappeared inside the
horizon, and is the result of the fact that at a fixed radius,
decreasing $\theta$ also decreases the flat space circular radius
$\rho=r\,\sin\theta$, which increases the centripetal acceleration
roughly by $1/\rho$ compared to the central attraction which does
not dramatically change with angle. This in turn increases the
parallel transport rotation-dominated zone compared to the
boost-dominated zone. The fact that the velocity profile
corresponding to $\tau_2=0$ (where $I_1=0$), which determines the
extremum of the acceleration polar angle $\chi$, is an extremely
small negative function at these scales (explaining why all the
$I_1$ curves appear to pass through the origin in Fig.~2) shows
that this point on the acceleration hyperbola is just slightly to
the counterrotating side of the ZAMO point.

In the case of the equatorial plane of the Kerr spacetime
discussed extensively in [\citen{bjm2}], the acceleration is
purely radial corresponding to $\chi=0,\,\pi$ which makes its
$\chi$-derivative and hence $\tau_2$ identically zero. Thus $A$ is
always semi-singular and becomes singular at the boundary between
boost and rotation dominance where $|\kappa| = |\tau_1|$. This
occurs for
\begin{equation}\label{eq:nuPTeq}
\nu_{(PT \pm)} = \frac{\nu_{\rm(gmp)}\mp \nu_+ \,\nu_- }{1\mp
\nu_{\rm(gmp)}} \ ,\quad \nu_{\rm(gmp)} = (\nu_++\nu_-)/2\ ,
\end{equation}
where $\nu_{\rm(gmp)}$ is the  velocity of the geodesic meeting
point trajectories \cite{bjm2}, or explicitly
\beq\label{eq:nuPTgmp} \nu_{(PT\pm)} =
\frac{m\left[a\,(r-m)\,(a^2+3\,r^2)\pm(r^3+a^2\,r+2\,a^2\,m)\,\sqrt{\Delta}\right]}
       {(-r^4+2\,m\,r^3 + 2\,a^2\,m\,r + a^2\, m^2)\,\sqrt{\Delta}} \ .
\eeq Notice that from the formula (\ref{eq:nuPTeq}), the values
the values $\nu_{(PT\pm)} = \pm1$ occur when respectively $\nu_\pm
=\pm1$, so that the respective endpoints of the boost-dominated
zone go null at the same radii as the counter-rotating ($r_{I}$)
and then the corotating circular geodesics ($r_{II}$) go null as
one approaches the horizon. Similarly, for example, as
$\nu_{\rm(gmp)}\to-1^+$, then $\nu_{\rm(PT-)}\to-\infty$, which
occurs at $r_{III}$ in Fig.~\ref{fig:6}. This feature of the
geodesic meeting point horizon in the equatorial plane was
uncovered in studies of circular holonomy started by Rothman et al
\cite{rot,MMM,bjm2}.

From  Eqs.~(4.9) and (4.10) of [\citen{idcf2}], the conditions
$\nu_\pm=\pm 1$ are equivalent to $\nu_{\rm (crit)\pm}=\pm 1$, so
in turn the conditions $\nu_{(PT -)} =-1 $ and $\nu_{(PT +)} =1$
correspond to the horizons $r_{I}$ and $r_{II}$ of the extremely
accelerated orbits in the equatorial plane, which are critical
points of the acceleration as a function of the relative velocity,
labeled as crit$\pm$ in Fig.~\ref{fig:6}. This explains the triple
crossing points of the geodesic velocities, extremely accelerated
velocities and the boost zone endpoint velocities in that figure.
The solutions of the conditions $\nu_{(PT\pm)} =\pm1 $, when
fractions are cleared (numerator equals plus or minus the
denominator) to produce polynomial equations in $r$ after some
manipulation to eliminate the radical, have a common factor
$4\,a^2\,m-9\,m^2\,r+6\,m\,r^2-r^3$, whose zeros produce the radii
$r_{I}$ and $r_{II}$ of the pair of null circular geodesics.  The
zero of the denominator of Eq.~(\ref{eq:nuPTgmp}) is instead
$r_{III}=r_{\rm(gmp)}$, the horizon of the geodesic meeting point
observers.

\begin{figure}[!t]
\typeout{Figure 9}
\centerline{\epsfxsize=0.48\textwidth \epsfbox{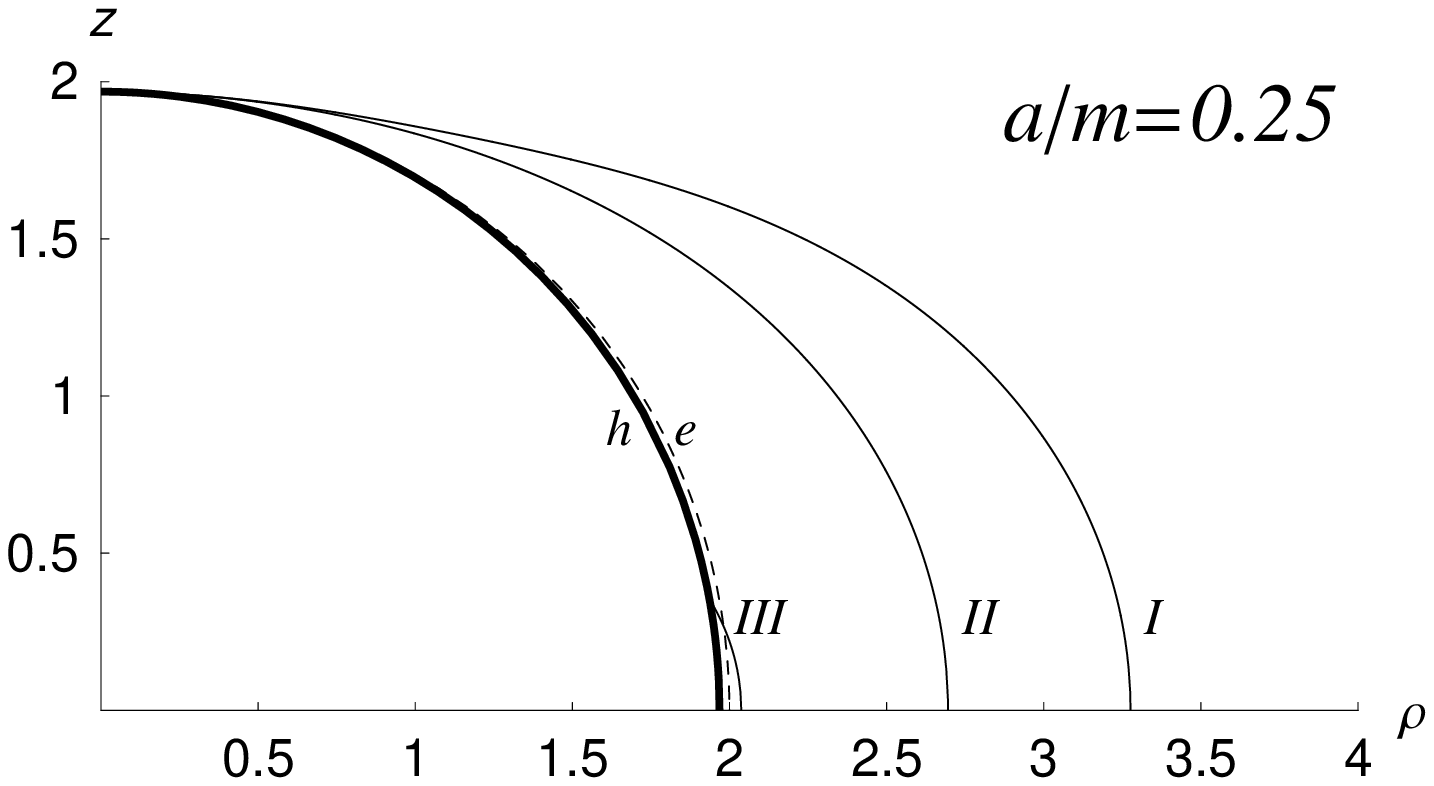}\hfil
\epsfxsize=0.48\textwidth\epsfbox{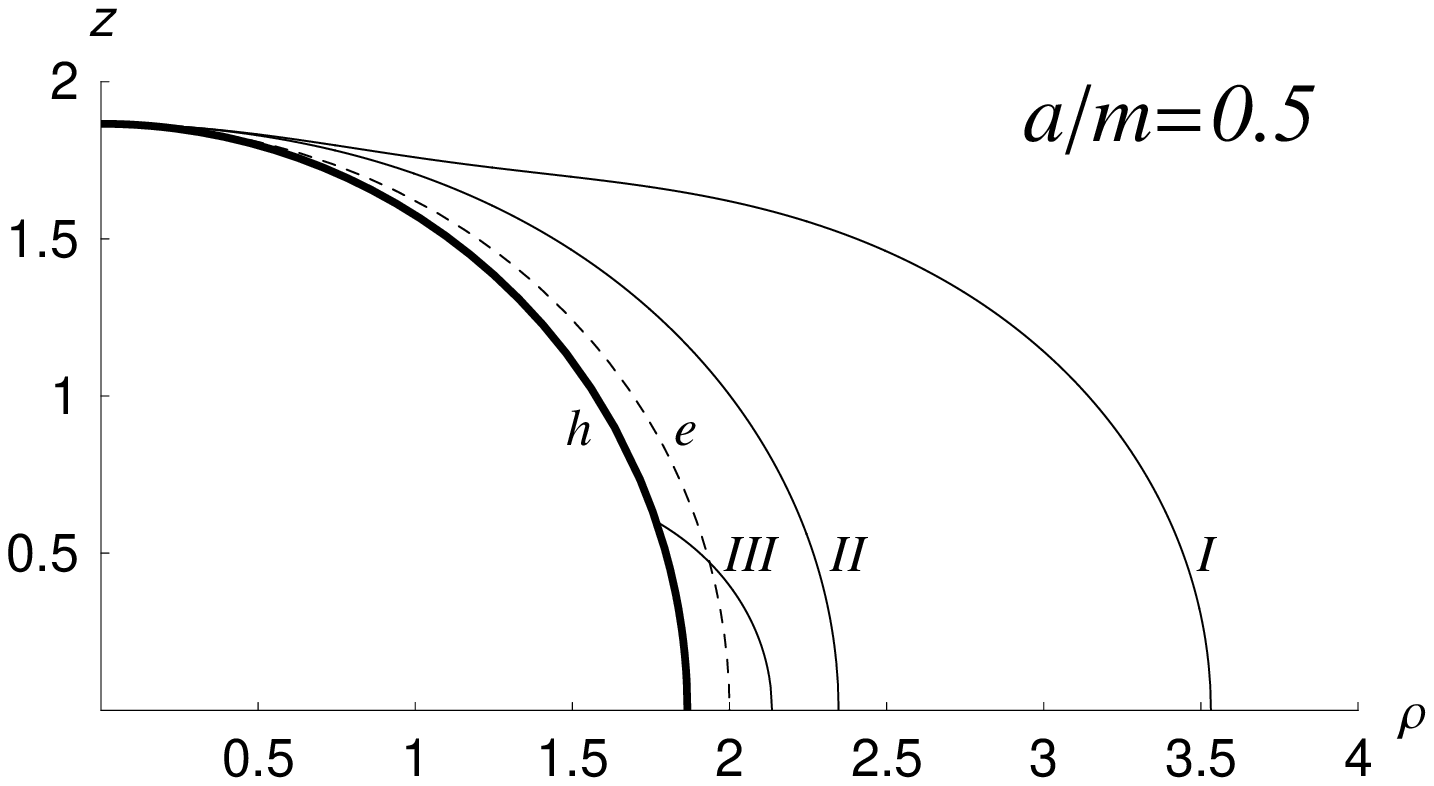}}
\centerline{\epsfxsize=0.48\textwidth \epsfbox{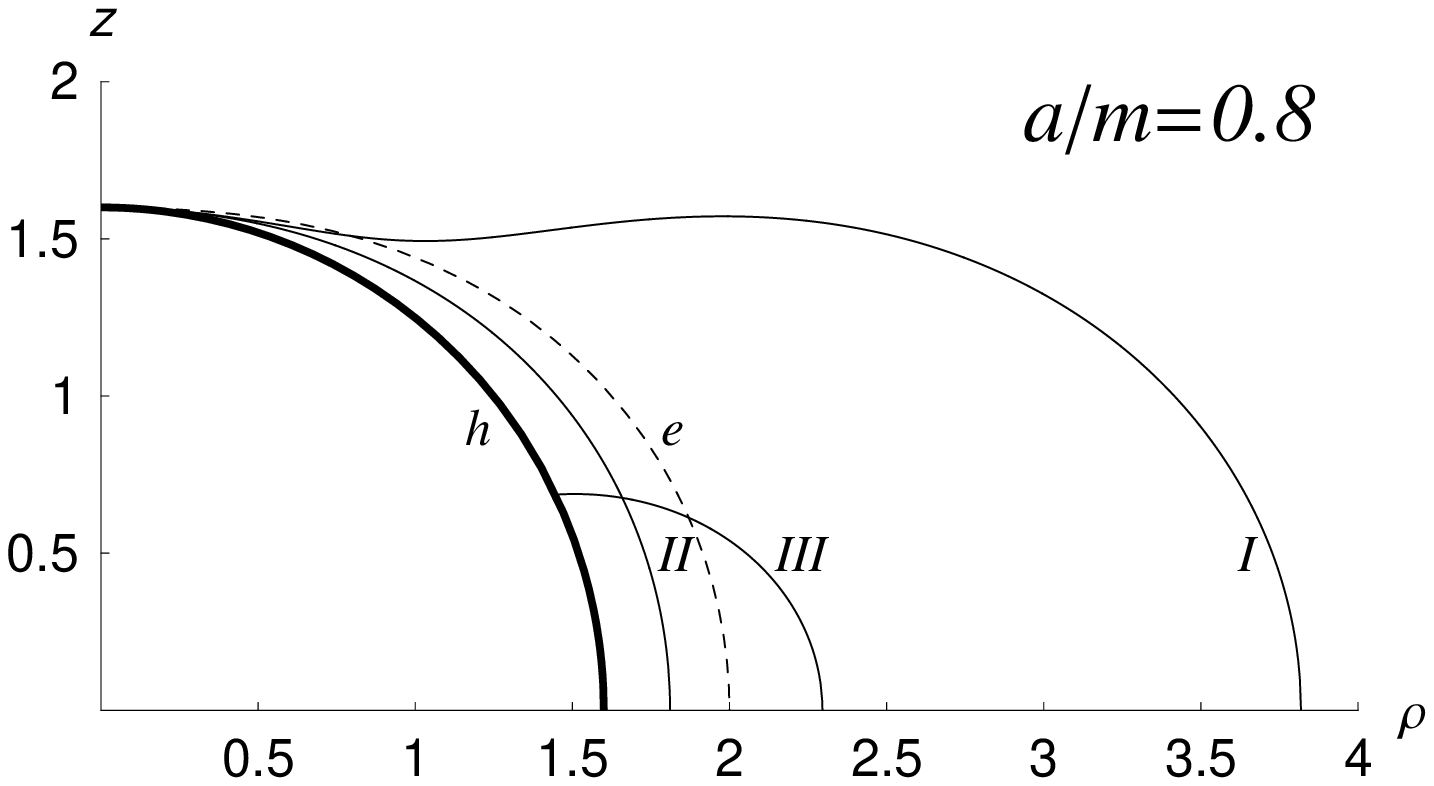}\hfil
\epsfxsize=0.48\textwidth \epsfbox{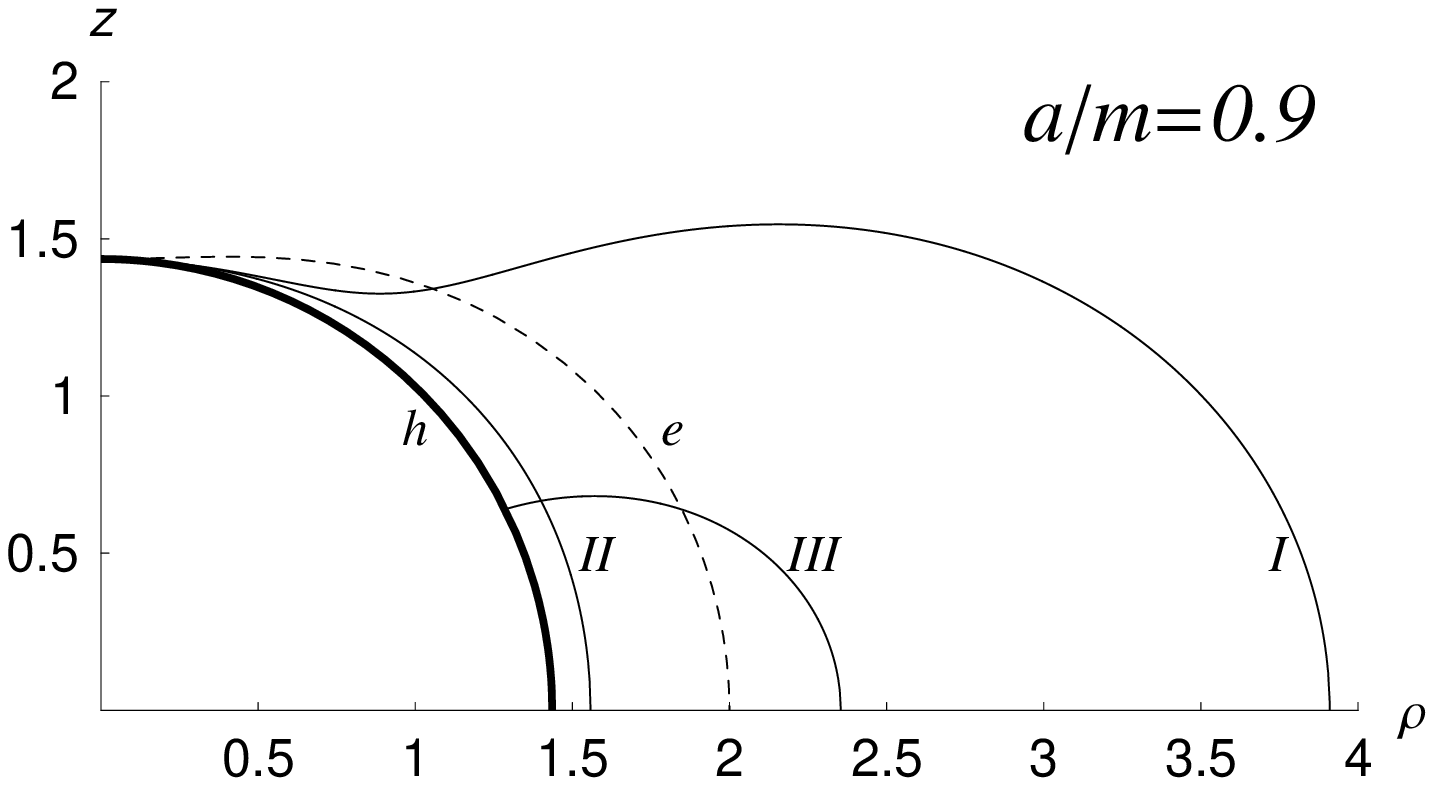}}
\centerline{\epsfxsize=0.48\textwidth \epsfbox{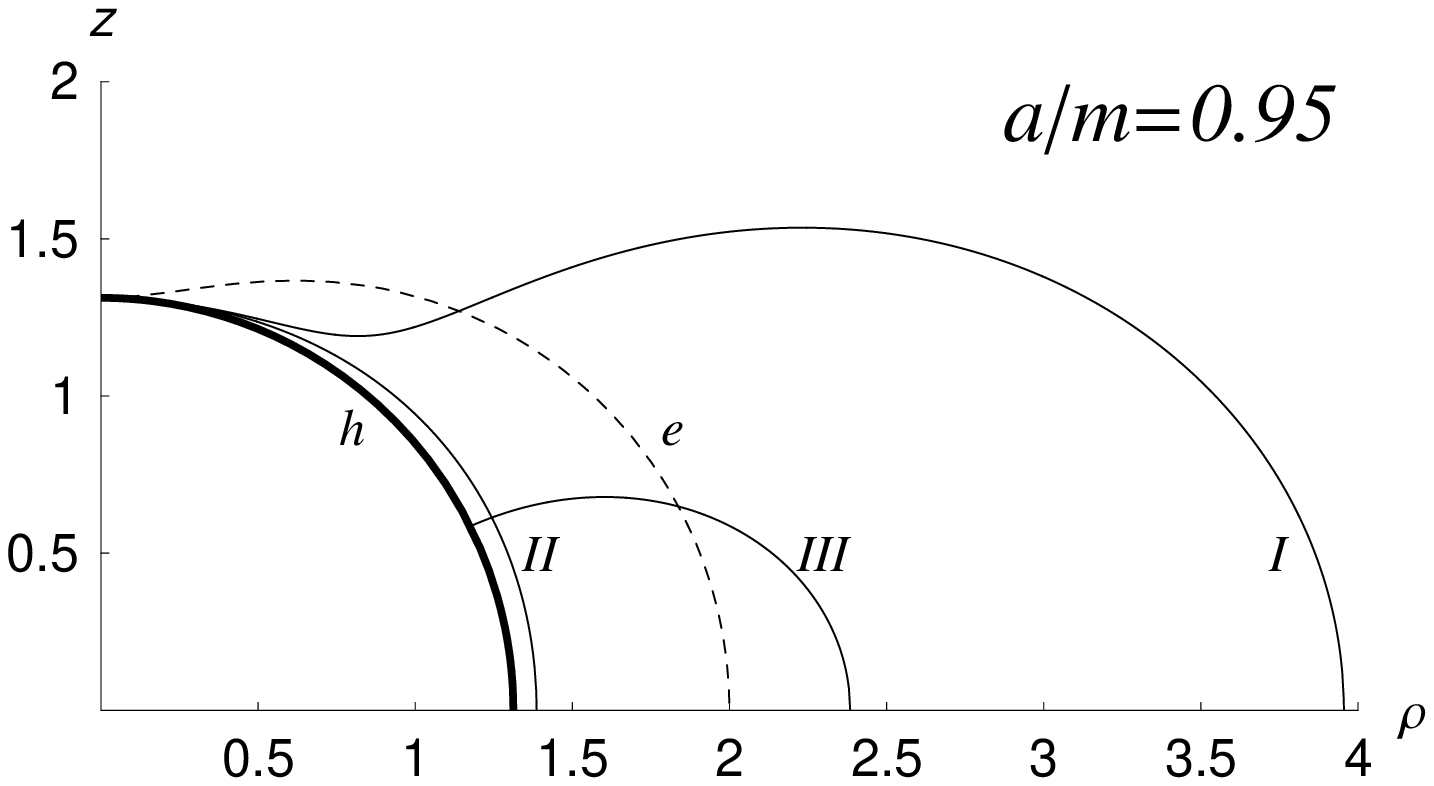}\hfil
\epsfxsize=0.48\textwidth \epsfbox{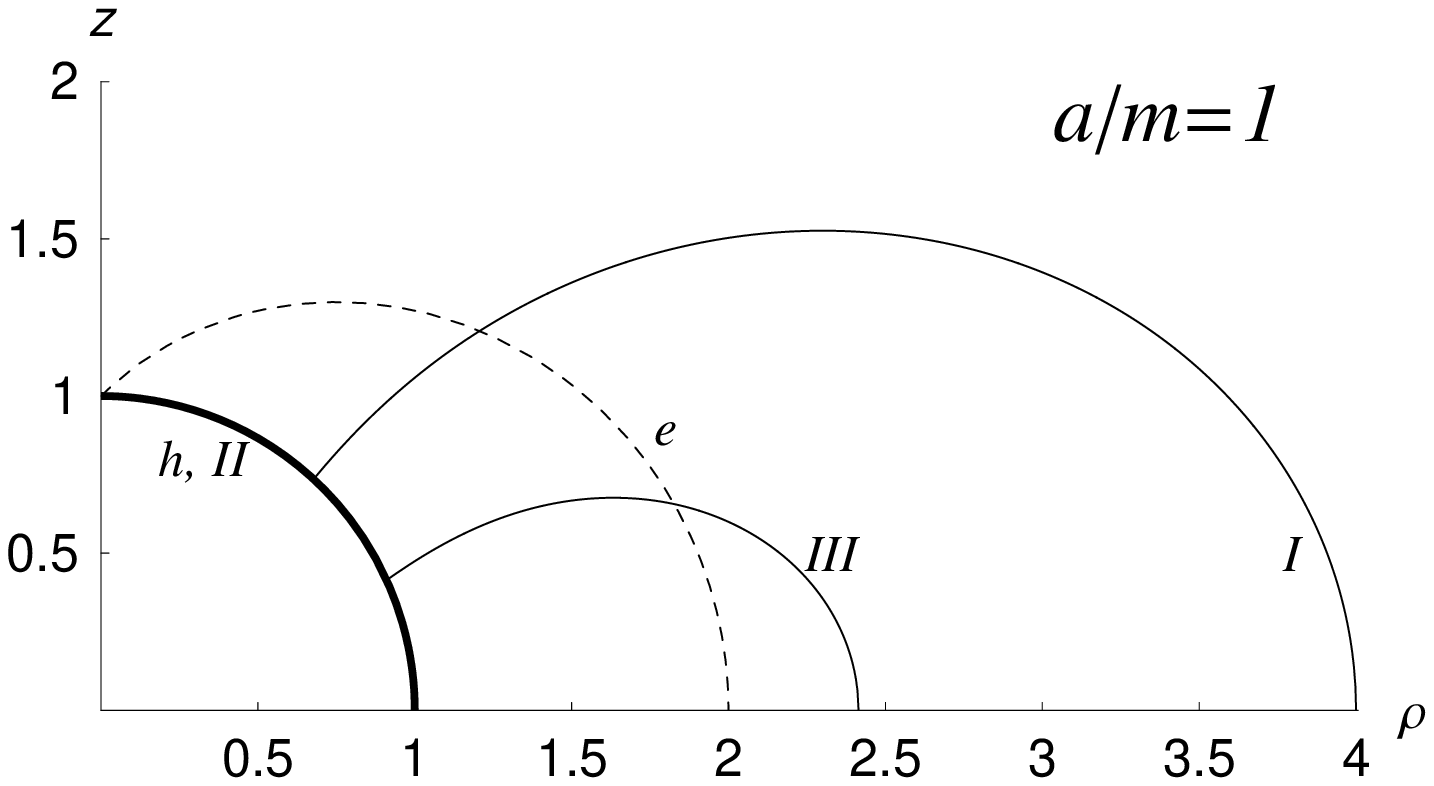}} \caption{ The
figure corresponding to Fig.~\ref{fig:7} as $a/m$ increases from
1/4 towards the extreme value 1. The surface $II$ collapses to the
horizon, while the surfaces $I$ and $III$ bubble upwards and
outwards (like the ergosphere boundary which they cross) to form
two doughnut-like rings (when revolved about the $z$-axis) around
the horizon, one inside the other. In the Schwarzschild limit
$a/m\to0$, the surfaces $I$ and $II$ collapse together to a single
surface meeting the equatorial plane at $r/m=3$, while the
ergosphere boundary and surface $III$ collapse together to the
horizon. } \label{fig:9}
\end{figure}

As $a/m$ increases from the value $1/2$ of Fig.~\ref{fig:6} to the
extreme limit $a/m=1$ in the equatorial plane, the radius $r_{II}$
shrinks down to the horizon  $r_{\rm(hor)}$ crossing over
$r_{III}$ as shown in Fig.~2 of \cite{bjm2} where these radii are
labeled respectively by ``geo$_+$," ``gmp" and ``hor." In fact
very interesting changes occur in the interval of values
$[0.89,1]$ for $a/m$ as illustrated in the polar plots of
Fig.~\ref{fig:9}. As the surface $II$ shrinks down to the horizon,
the surfaces $III$ and $I$ both bubble upwards and outwards, with
surface $I$ developing a transversal intersection with the horizon
only at the limiting value $a/m=1$ of an extreme Kerr spacetime.
That surface $II$ shrinks to the horizon in this limit means that
there always exists a rotation-dominated zone for sufficiently
fast but subluminal corotating orbits outside the horizon.  The
increasing separation between the horizon and surface $III$ away
from the North pole in this limit means that the corresponding
small piece of grey boost-dominated region at the top left corner
of Figs.~\ref{fig:6} and \ref{fig:8}$a)$ grows outward and
downward as the lower grey zone falls downward following the
rotation-dominated region as its lower border in turn moves
downwards. Simultaneously the counterrotating boost-dominated
region moves outward from the horizon away from the North pole.
Thus the faster the spacetime rotates, the worse things are for
the counterrotating orbits, which go superluminal farther and
farther away from the shrinking horizon leading to
boost-domination inside that radius for all subluminal motion. At
the same time the corotating orbits go superluminal closer and
closer to the horizon, with expanding rotating-dominated zones
existing for suitably fast but subluminal motion outside that
radius.

\section{Concluding remarks}

The gravitoelectromagnetic analysis of parallel transport of the
4-velocity along generic circular orbits in stationary
axisymmetric spacetimes has been extended to parallel transport of
the full tangent space and connected to the corresponding
Serret-Frenet description. The electromagnetic decomposition of
the parallel transport matrix is essential to putting it into a
canonical form for interpretation. Crucial features of parallel
transport take place at the surfaces where various geometrically
defined 4-velocities become null. These results enable the
previous discussion of the equatorial plane of the Kerr spacetime
to be extended to the whole exterior spacetime outside the
horizon.

One finds new surfaces surrounding a black hole which
intrinsically characterize limiting parallel transport properties
along circular orbits of all causality types. Within these
surfaces, one has an invariant characterization of the domination
of centripetal acceleration effects by attractive forces,
differing of course for corotating and counterrotating orbits.
Examining the way in which the gravitoelectric, gravitomagnetic
and space curvature effects enter into the parallel transport
matrix and the resulting 4-acceleration gives some more insight
into how centripetal acceleration and gravitational attraction
compete with each other and are in turn affected by
gravitomagnetic effects in the Kerr spacetime, thus complementing
the results of previous work.\cite{circfs}

\appendix

\section{Geometry of the acceleration and Frenet-Serret angular velocity curves}

Using the defining condition for the MIROs (vertex observers),
(see the paragraph containing equation (6.3) of [\citen{circfs}])
$
e^{2\alpha_{\rm (vert)}}
 = \sqrt{||\vec a_-|| / ||\vec a_+||}
$, and introducing the quantity ${\cal A} = \sqrt{||\vec a_-||
||\vec a_+||}/2 $, the definitions of section 4 together with the
decomposition $a_\pm =||\vec a_\pm||\, \vec n_\pm $ into lengths
and unit direction vectors lead to
\begin{eqnarray}
\fl
(\vec a -\vec a_0)
 &=&
    \frac14 \left[ ||\vec a_+\,||e^{2\alpha}\,\vec n_+
     + ||\vec a_-||\,e^{-2\alpha}\,\vec n_- \right]
\nonumber\\
\fl
 &=& \frac{\cal A}{2}\,
    [e^{2(\alpha -\alpha_{\rm (vert)})}\,\vec n_+  +
            e^{-2(\alpha -\alpha_{\rm (vert)})}\,\vec n_-]
\\
\fl
 &=& \frac{\cal A}{2}\,
     [\cosh 2(\alpha -\alpha_{\rm (vert)}) \,(\vec n_+ + \vec n_-)
         +\sinh 2(\alpha -\alpha_{\rm (vert)})\, (\vec n_+ - \vec n_-)]
\ .\nonumber
\label{acc3}
\end{eqnarray}
Here $\alpha=\mbox{arctanh}\,\nu$ is the rapidity parametrization
of the acceleration curve.

The photon acceleration direction vectors,
which determine the common asymptotes of the acceleration and
Frenet-Serret angular velocity hyperbolas,
are independent of the reference observer family used to decompose the acceleration.
It is natural to introduce the unit direction vectors of their sum and difference which
determine the common orthogonal axes of the acceleration hyperbola.

The opening angle $\beta\in [0,\,\pi]$ of the acceleration
hyperbola (making $\pi-\beta$ the opening angle of the
Frenet-Serret angular velocity  hyperbola) satisfies $ \vec n_+
\cdot \vec n_- =\cos\beta $, in terms of which one finds
\begin{equation}
||\vec n_+ +\vec n_-|| /2
  = \cos \frac{\beta}{2}\nonumber
\ ,\quad
||\vec n_+ -\vec n_-|| /2
 = \sin \frac{\beta}{2} \ ,
\end{equation}
leading to the orthogonal unit vectors \beq \vec {\mbox{\i}} =
\frac{\vec n_+ + \vec n_-}{2\,\cos\frac{\beta}{2} } \ ,\qquad
\vec{\mbox{\j}} = \frac{\vec n_+ - \vec
n_-}{2\,\sin\frac{\beta}{2}} \eeq which give the directions of the
principal axes of both hyperbolas. The quantities ${\cal A}\,\cos
\frac{\beta}{2}$ and ${\cal A}\,\sin \frac{\beta}{2}$ are the
semi-axis parameter lengths of the acceleration and the
Frenet-Serret angular velocity hyperbolas. By introducing the sign
$\epsilon$ which makes the following cross-product relation valid
\begin{equation}\label{eq:ixj}
  \epsilon\, \vec{\mbox{\i}}\times_n \vec{\mbox{\j}}
=  e_{\hat r} \times_n e_{\hat\theta} = e_{\hat \phi} \ ,
\end{equation}
and whose value in the Kerr spacetime is $  \epsilon={\rm sgn}(\pi/2-\theta)$,
one can finally write the equations for the acceleration and angular velocity hyperbolas in terms of their geometric properties
\begin{eqnarray}
\fl \vec a -\vec a_0 = {\cal A}\,[\cosh 2(\alpha -\alpha_{\rm
(vert)})\, \cos \frac{\beta}{2} \;\vec{\mbox{\i}}
 + \sinh 2(\alpha -\alpha_{\rm (vert)})\, \sin \frac{\beta}{2} \;\vec{\mbox{\j}} ]\ ,
\nonumber\\
\fl \epsilon\, \vec\omega_{\rm(FS)} = {\cal A}\,
 [\cosh 2(\alpha -\alpha_{\rm (vert)})\,  \sin \frac{\beta}{2} \,\vec{\mbox{\i}}
 -\sinh 2(\alpha -\alpha_{\rm (vert)})\, \cos \frac{\beta}{2}\,\vec{\mbox{\j}}
 ]\ .
\end{eqnarray}

It is then easy to derive the following formulas for the
acceleration and angular velocity scalars
\begin{eqnarray}
2 (\vec a -\vec a_0) \cdot \vec\omega_{\rm(FS)}
&=&
{\cal A}^2 \sin\beta \ ,
\nonumber\\
||\vec a -\vec a_0||^2 &=& \frac12\, {\cal A}^2 [\cosh 4(\alpha
-\alpha_{\rm (vert)})+\cos\beta]\ ,
\nonumber\\
||\vec\omega_{\rm(FS)}||^2 &=& \frac12\, {\cal A}^2 [\cosh
4(\alpha -\alpha_{\rm (vert)}) - \cos\beta] \ , \label{eq:omegaeq}
\end{eqnarray}
or equivalently
\begin{eqnarray}
||\vec a -\vec a_0||^2- ||\vec\omega_{\rm(FS)}||^2
&=&
{\cal A}^2 \cos\beta \ ,
\nonumber \\
||\vec a -\vec a_0||^2+ ||\vec\omega_{\rm(FS)}||^2
&=&
{\cal A}^2
    \cosh 4\,(\alpha -\alpha_{\rm (vert)})\ .
\end{eqnarray}


\section{Frenet-Serret frame for circular orbits}

The sign conventions for the Frenet-Serret scalars and the
corresponding directional choices of the frame vectors for nonnull
circular orbits in stationary axisymmetric spacetimes require
careful consideration \cite{circfs}. Starting from the rapidity
parametrized 4-velocity for a future-pointing timelike circular
orbit \beq\label{eq:B1}
  U = \cosh\alpha\, n + \sinh\alpha\, e_{\hat\phi} = e_0
\eeq expressed in a Boyer-Lindquist-like ZAMO orthonormal frame
$\{n,\,e_{\hat r},\,e_{\hat\theta},\,e_{\hat\phi}\}$, one can
introduce a unit vector in the direction of increasing $\phi$ in
the local rest space of $U$, namely \beq
 e_2 = \frac{\partial}{\partial\alpha}\, U
     = \sinh\alpha\, n + \cosh\alpha\, e_{\hat\phi} \ ,
\eeq thus defining two Frenet-Serret vectors in the circular orbit
velocity plane so that $e_0\wedge e_2 = n\wedge e_{\hat\phi}$. It
remains to be seen that $e_2$ is actually a possible choice.

Next one introduces polar coordinates in the acceleration plane
spanned by $\{e_{\hat r},\,e_{\hat\theta}\}$ to parametrize the
acceleration vector, with a nonnegative radial coordinate \beq
  a(U) = \kappa\, ( \cos\chi\, e_{\hat r} + \sin\chi\, e_{\hat\theta})
    = \kappa\, e_1
  \ ,\quad \kappa\geq0\ ,
\eeq
so that $e_1$ is the unit direction of $a(U)$ and $e_3$ is the orthogonal unit vector
\beq
 e_3 = -\frac{\partial}{\partial\chi}\, e_1
     = \sin\chi\, e_{\hat r} - \cos\chi\, e_{\hat\theta} \ ,
\eeq
satisfying
\beq
e_{\hat \phi}\times_n e_1 = -e_3\ , \quad
e_{\hat \phi}\times_n e_3 = e_1\ , \quad
e_3\wedge e_1 = e_{\hat r} \wedge e_{\hat\theta}\ ,
\eeq
and so
$e_0\wedge e_1 \wedge e_2 \wedge e_3
  = n \wedge e_{\hat r} \wedge e_{\hat\theta} \wedge e_{\hat\phi}$.
Since $n$ and $e_0$ both future-pointing timelike vectors, the
Frenet-Serret spatial triad is also right-handed.

To show that $e_2$ and $e_3$ can consistently be chosen as the remaining Frenet-Serret vectors, the transformation (\ref{eq:EBUgH}) of the connection matrix $A$ must be used together with the identifications (\ref{eq:EBtauomega}), leading to
\begin{eqnarray}
\pmatrix{ -\kappa\, e_1\cr \tau_2\,e_1 + \tau_1\,e_3\cr } &=&
M(2\alpha) \pmatrix{ g_- \cr H\cr} + \pmatrix{ g_+ \cr 0 \cr}
\nonumber\\
&=&
\pmatrix{
\cosh 2\alpha \, g_- +\sinh 2\alpha \, e_{\hat \phi}\times_n H + g_+ \cr
-\sinh 2\alpha \, e_{\hat\phi}\times_n g_- +\cosh 2\alpha\,   H \cr
}
\ .
\end{eqnarray}
Then since $g_+ $ is independent of $\alpha$ and $e_{\hat
\phi}\times_n (e_{\hat \phi}\times_n H)=-H$, the Frenet-Serret
angular velocity can be rewritten as
\begin{eqnarray}
\tau_2\,e_1+\tau_1\,e_3 &=& \frac12\, e_{\hat \phi}\times_n
\frac{d}{d\alpha } (\kappa\, e_1) = \frac12\,
\frac{\partial\kappa}{\partial\alpha}\, e_{\hat \phi}\times_n e_1
 +\frac12\, \kappa \,\frac{\partial\chi}{\partial\alpha}\,
        e_{\hat \phi}\times_n \frac{\partial e_1}{\partial\chi}
\nonumber\\
&=& - \frac12 \,\frac{\partial\kappa}{\partial\alpha}\, e_3
    - \frac12\, \kappa\, \frac{\partial\chi}{\partial\alpha}\, e_1\ .
\end{eqnarray}
Thus everything is consistent if one makes the identification
\beq\label{eq:torsions} \tau_1 = -\frac12
\,\frac{d\kappa}{d\alpha}\ , \qquad \tau_2 = -\frac12\, \kappa\,
\frac{d\chi}{d\alpha}\ . \eeq The Frenet-Serret angular velocity
is then given by the rapidity derivative formula
\beq\label{eq:omegaFS} \omega_{\rm(FS)} = \frac12\,
e_{\hat\phi}\times_n
                \frac{\rmd a(U)}{\rmd\alpha}
= \cosh2\alpha \, H - \sinh2\alpha\, e_{\hat\phi}\times_n  g_-
\ .
\eeq

Note that interchanging the hyperbolic functions in (\ref{eq:B1})
leads to the corotating spacelike case for $U$, which only
interchanges $U$ and $e_2$, resulting in the same identifications
(\ref{eq:torsions}) and the same first equation of
(\ref{eq:omegaFS}). It remains to check all these signs for the
remaining two cases of the four sign combinations for
parametrizing $U$ by a hyperbolic angle $\alpha$ as in
Eq.~(\ref{eq:Ualpha}).


\end{document}